\documentclass[aps,pre,twocolumn,amsfonts,amsmath,showpacs,a4paper,floatfix]{revtex4}
\usepackage{enumerate}
\usepackage{amsfonts}
\usepackage{amssymb}
\usepackage{amsmath}
\usepackage{amscd}
\usepackage{color}
\usepackage{psfrag}
\usepackage{latexsym}
\usepackage[dvips]{graphicx}
\usepackage{amsthm}

\newcommand{\eqlabel}[1]{\label{eq:#1}}
\newcommand{\eq}[1]{(\ref{eq:#1})}

\newcommand{\dblfigure}[3]
  {\begin{figure*}[tbp]\centerline{\includegraphics{#1}}\caption[]{#2}\figlabel{#3}\end{figure*}}
\newcommand{\sglfigure}[3]
  {\begin{figure}[tbp]\centerline{\includegraphics{#1}}\caption[]{#2}\figlabel{#3}\end{figure}}

\newcommand{\figlabel}[1]{\label{fig:#1}}
\newcommand{\figref}[1]{\ref{fig:#1}}
\newcommand{\fig}[1]{fig.~\figref{#1}}
\newcommand{\figs}[1]{figures~\figref{#1}}
\newcommand{\Fig}[1]{Figure~\figref{#1}}

\newcommand{\seclabel}[1]{\label{sec:#1}}
\newcommand{\secref}[1]{\ref{sec:#1}}
\newcommand{\secn}[1]{section~\secref{#1}}

\newcommand{\apriori}{\textit{a~priori}}
\newcommand{\eg}[1]
  {{\it e.g.\/}\ifx#1.\else\expandafter#1\fi}
\newcommand{\etal}[1]
  {{\it et al.\/}\ifx#1.\else\expandafter#1\fi}
\newcommand{\ibid}[1]
  {{\it ibid.\/}\ifx#1.\else\expandafter#1\fi}
\newcommand{\ie}[1]
  {{\it i.e.\/}\ifx#1.\else\expandafter#1\fi}

\renewcommand{\d}{\mathrm{d}}           
\newcommand{\Df}[2]
  {\frac{\d{#1}}{\d{#2}}}
\newcommand{\df}[2]
  {\frac{\partial{#1}}{\partial{#2}}}

\newcommand{\Mx}[1]{\left[\begin{array}{cccccc} #1 \end{array}\right]} 
\newcommand{\mx}[1]{\mathbf{#1}}        

\newcommand{\A}{\mx{A}}                 
\newcommand{\Acal}{\mathcal{A}}         
\newcommand{\Ahat}{\hat\Acal}           
\newcommand{\bzero}{\mathbf{0}}         
\newcommand{\B}{{\cal{B}}}              
\renewcommand{\c}{{\vec c}}             
\newcommand{\chat}{\hat\c}              
\newcommand{\codim}{\mathop{\mathrm{codim}}} 
\newcommand{\D}{\mx{D}}                 
\newcommand{\dM}{\partial M}            
\newcommand{\dR}{\partial R}            
\newcommand{\E}{\mx{E}}                 
\newcommand{\F}{\mx{F}}                 
\newcommand{\Fg}{\F_{\G}}               
\newcommand{\Fm}{\F_{\M}}               
\newcommand{\Fcal}{\mathcal{F}}         
\newcommand{\Fhat}{\hat\Fcal}           
\newcommand{\f}{\mx{f}}                 
\newcommand{\G}{\mathcal{G}}            
\newcommand{\g}{g}                      
\newcommand{\gi}{\g^{-1}}               
\renewcommand{\H}{\mx{H}}               
\newcommand{\Htil}{\tilde{\H}}          
\newcommand{\Hg}{\mx{H}_{\G}}           
\newcommand{\Hm}{\mx{H}_{\M}}           
\newcommand{\Hcal}{\mathcal{H}}         
\newcommand{\Hhat}{\hat\Hcal}           
\newcommand{\h}{\mx{h}}                 
\newcommand{\htil}{\tilde{\h}}          
\newcommand{\id}{\mathrm{id}}           
\newcommand{\iinc}{i_{\textrm{inc}}}    
\newcommand{\jinc}{j_{\textrm{inc}}}    
\renewcommand{\L}{L}                    
\newcommand{\M}{\mathcal{M}}            
\newcommand{\mxi}{\hat{\boldsymbol{\gamma}}} 
\newcommand{\Nx}{N_x}                   
\newcommand{\Ny}{N_y}                   
\newcommand{\omegahat}{\hat\omega}      
\newcommand{\Real}{\mathbb{R}}          
\newcommand{\SEtwo}{SE(2)}              
\newcommand{\R}{{\vec R}}               
\newcommand{\Rhat}{\hat\R}              
\renewcommand{\r}{{\vec r}}             
\newcommand{\rinc}{\r_{\textrm{inc}}}   
\renewcommand{\S}{\mathcal{S}}          
\newcommand{\stept}{\Delta_t}           
\newcommand{\stepx}{\Delta_x}           
\newcommand{\T}{\mathrm{T}}             
\newcommand{\Thetahat}{\hat\Theta}      
\newcommand{\U}{\mx{U}}                 
\renewcommand{\u}{\mx{u}}               
\newcommand{\V}{\mx{V}}                 
\newcommand{\Vhat}{\hat\V}              
\renewcommand{\v}{\mx{v}}               
\newcommand{\vhat}{\hat{\v}}            
\newcommand{\vzero}{\vec{0}}            
\newcommand{\ustar}{u_*}                
\newcommand{\vstar}{v_*}                
\newcommand{\xtip}{x_{\textrm{tip}}}    
\newcommand{\ytip}{y_{\textrm{tip}}}    

\newcommand{\ezspiral}{EZ-SPIRAL}
\newcommand{\ezride}{EZRide}

\begin{document}

\title{
  Riding a Spiral Wave: \\ 
  Numerical Simulation of Spiral Waves in a Co-Moving Frame of Reference
}

\author{A.J.~Foulkes}
\affiliation{Department of Computer Science, University of Liverpool, 
     Ashton Building, Ashton Street, Liverpool L69 3BX, UK}

\author{V.N.~Biktashev}
\affiliation{Department of Mathematical Sciences, University of Liverpool, 
   Mathematical Sciences Building, Peach Street, Liverpool, L69 7ZL, UK}
   
\date{\today}

\begin{abstract}
We describe an approach to numerical simulation of spiral waves dynamics 
of large spatial extent, using small computational grids. 
\end{abstract}

\pacs{%
  02.70.-c, 
  05.10.-a, 
  82.40.Bj, 
  82.40.Ck, 
  87.10.-e  
}

\maketitle


\section{Introduction}\seclabel{introduction}

Spiral waves are a type of self-organization observed in a large
variety of spatially extended, thermodynamically non-equilibrium
systems of physical, chemical and biological nature~\cite{
  Zhabotinsky-Zaikin-1971,%
  Allessie-etal-1973,%
  Alcantara-Monk-1974,%
  Carey-etal-1978,%
  Gorelova-Bures-1983,%
  Schulman-Seiden-1986,%
  Murray-etal-1986,%
  Madore-Freedman-1987,%
  Jakubith-etal-1990,%
  Lechleiter-etal-1991,%
  Frisch-etal-1994,%
  Shagalov-1997,%
  Yu-etal-1999,%
  Agladze-Steinbock-2000,%
  Dahlem-Mueller-2003,%
  Igoshin-etal-2004,%
  Larionova-etal-2005,%
  Oswald-Dequidt-2008,%
  Bretschneider-etal-2009%
}, where wave propagation is supported by a source of energy stored in
the medium.  If the system can be considered spatially uniform and
isotropic and its properties do not depend on time, the corresponding
mathematical models possess corresponding symmetries.  For many
practical applications, a considerable interest is in non-stationary
dynamics of spiral waves, which is usually defined separately either
as drift, which is displacement of the average position of the core of
the spiral with time due to external symmetry-breaking perturbations,
or meandering, which is spontaneous symmetry breaking due to internal
instability rather than external forces and which is
manifested by
complicated movement of the spiral with the average position of the
core typically unmoved.

The numerical simulation of drift and meander of spiral waves,
particularly when models are complicated and high accuracy is
required, can be challenging. There are some theoretical
considerations which suggest some way of dealing with this challenge.
So it has been observed that as far as drift is concerned, spiral
waves behave like particle-like objects, which results from effective
localization of the critical eigenfunctions of the adjoint linearized
operator~\cite{%
  Biktashev-1989,%
  Biktashev-etal-1994,%
  Biktashev-Holden-1995,%
  Biktasheva-etal-1998,%
  Biktasheva-Biktashev-2003%
}, so it should be sufficient to do the computations only around the
core of the spiral to predict its drift.  On the other hand, in the absence
of external symmetry breaking perturbations, meandering of spirals can
be understood by explicitly referring to the Euclidean symmetry of the
unperturbed problem~\cite{%
  Barkley-1994,%
  Barkley-Kevrekidis-1994,%
  Barkley-1995,%
  Mantel-Barkley-1996,%
  Biktashev-etal-1996,%
  Wulff-1996%
}. Specifically, an idea of dynamics in the space of symmetry group
orbits~\cite{Chossat-2002}, when applied to a reaction-diffusion
system of equations and the Euclidean symmetry group, leads to a
description which is formally equivalent to considering the solution
in a moving frame of reference (FoR) such that the spiral wave
maintains a certain position and orientation in this
frame~\cite{Biktashev-etal-1996}. We shall call it comoving FoR for
short.

The purpose of this article is to present a computational approach
based on these considerations. We calculate the dynamics of the spiral
wave in a comoving FoR; as a result, the core of the spiral never
approaches the boundaries of the computation box, which allows
computations of drift and meandering of large spatial extent using
small numerical grids. A simple software implementation of this
approach, which is based on the popular spiral wave simulator
`\ezspiral'~\cite{Barkley-1991,EZSpiral}, and which we called
'\ezride', is provided on the authors' website~\cite{EZRide}.

Our approach can be compared to the approach proposed by Beyn
  and Thummler~\cite{Beyn-Thummler-2004} and further developed by
  Hermann and Gottwald~\cite{Hermann-Gottwald-2010}. Their approach also exploits symmetry group orbits,
  but is different in some essential details. We shall discuss
  the similarities and differences when we will have introduced our
  method.

The structure of the paper is as follows. In \secn{maths} we lay out
mathematical basics of the approach~and briefly
    compare it with \cite{Beyn-Thummler-2004}.
In \secn{numerics} we describe
the numerical method itself. In \secn{examples} we illustrate the work
of the method by simple and quick examples. The potential for
numerical accuracy is demonstrated in \secn{convergence}.  The
subsequent three section are dedicated to examples of applications of
the methods to problems where the conventional methods would be
struggling: \secn{resonance} for the degenerate case of meandering
which results in ``spontaneous drift'' of spirals; \secn{largecore}
for the dynamics near to, and beyond, the parametric boundary at which
the core radius of the spiral becomes infinite; and \secn{drift} for
drift caused by a symmetry breaking perturbation.  We conclude with a
brief discussion of the results in
\secn{discussion}.


\section{Symmetry group reduction}\seclabel{maths}

Following \cite{Biktashev-etal-1996}, we start from a perturbed
reaction-diffusion system of equations in a plane,
\begin{equation}
\eqlabel{rdeu}
\df{\u}{t} = \D\nabla^2\u +\f(\u)+\h(\u,\nabla\u,\r,t),
\end{equation}
where $\u=\left(u^{(1)},\dots,u^{(n)}\right)^\top=\u(\r,t)\in\Real^n$
is a column-vector of reagent concentrations varying in space and
time, $\f=\f(\u)$ is a column-vector of reaction rates,
$\D\in\Real^{n\times n}$ is the matrix of diffusion coefficients,
$\h\in\Real^n$ represents symmetry-breaking perturbations,
$||\h||\ll1$, $n\ge 2$, and $\r=(x,y)\in\Real^2$. If
$\h=\bzero$, then equation~\eq{rdeu} is equivariant
with respect to Euclidean transformations of the
spatial coordinates $\r$.

The following technical discussion is necessary
to place our method in the context of other works in the field. 
Readers not interested in technical details may skip down to
system \eq{ezfnum}.

The idea of the symmetry group reduction is
convenient to describe if we view \eq{rdeu} as an ordinary
differential equation in a suitably chosen functional space
$\B$, 
\begin{equation}
 \Df{\U}{t} = \F(\U)+\H(\U,t)                       \eqlabel{rdeU}
\end{equation}
where $\U:\Real\to\B$ represents the dynamic field $\u$,
$\F:\B\to\B$ represents
the unperturbed right-hand side $\D\nabla^2\u+\f$, and
$\H:\B\times\Real\to\B$ represents the perturbation
$\h$.

Let us suppose that equation \eq{rdeU} at $\h=\bzero$ is equivariant
with respect to a representation $\T$ of a Lie group $\G$ in $\B$.
This means that for any $\g\in\G$ and any $\U\in\B$, we have 
\begin{equation}
  \F(\T(\g)\U) = \T(\g) \F(\U). \eqlabel{equivariat}
\end{equation}
In our case, $\G=\SEtwo$, the special Euclidean transformations of
the plane $\Real^2\to\Real^2$ (including translations and rotations),
and $\T$ is its representation in the space of functions $\u(\r)$
defined on this plane, acting as 
\begin{equation}
  \T(\g)\u(\r)=\u(\g^{-1}\r) .                             \eqlabel{T}
\end{equation}

We consider a subset $\B_0\subset\B$ such that $\G$ acts freely on
$\B_0$, \ie\ for a $\U\in\B_0$, any nontrivial transformation $\g\in\G$
changes $\U$, in other words, $\T(\g)\U=\U \;\Rightarrow\; \g=\id$. In
the terminology of \cite{Chossat-2002}, $\B_0$ is the principal stratum of
$\B$, corresponding to the trivial isotropy subgroups. In our case,
this means that the graph of the function $\u(\r)\in\B_0$ is devoid of
any rotational or translational symmetry, which is of course true for
functions describing single-armed spiral waves.

It is straightforward that at $\H=\bzero$, the set $\B_0$ is an
invariant set of \eq{rdeU}. Moreover, we shall restrict our
consideration to such perturbations $\H(t)$ that resulting
solutions $\U(t)$ remain in $\B_0$ for all $t$. This means, that the
perturbations are supposed to be so small they cannot impose
incidental symmetry on the otherwise unsymmetric spiral wave
solutions.

A group orbit of a given $\U$ is defined as the set
$\T(\G)\U=\{\T(\g)\U\,|\,\g\in\G\}$. That is, it is a set of all
such functions $\u(\r)$ that can be obtained from one another by
applying an appropriate Euclidean transformation to $\r$. 
A group orbit is a manifold in $\B_0$, 
of a dimensionality equal to $d=\dim\G$ 
less the dimensionality of the isotropy group.
In our case, $\dim\SEtwo=3$, the isotropy group is trivial and
the orbits are smooth three-dimensional manifolds. 

From the definition of the set $\B_0$ it follows that this set is 
foliated by
group orbits. The principal assumption for the following
analysis is that there exists an open subset $\S\subset\B_0$, also
invariant with respect to $\G$, in which the foliation has a global
transversal section, \ie\ we can select one representative from each
orbit in $\S$, such that all such representatives form a smooth
manifold $\M\subset\S$, which is everywhere transversal to the
group orbits. We call this manifold a \emph{Representative Manifold}
(RM). That would mean that any orbit in $\S$ crosses $\M$
transversally and exactly once. Hence
\begin{equation}
  \eqlabel{manifold}
  \forall\,\U\in \S,\quad\exists'\,(\g,\V)\in\G\times\M:\quad\U=\T(\g)\V.
\end{equation}
The RM has co-dimensionality equal to the dimensionality of the group
orbits, \ie\ in our case $\codim\M=d=3$.  It is assumed to be smooth
and we expect that it can locally be described by equations
$\mu_\ell(\V)=0$, $\ell=1,\dots d$, where functions
$\mu_\ell:\B\to\Real$, \ie\ are functionals when interpreted in terms
of the original reaction-diffusion equation \eq{rdeu}.

A convenient pictorial interpretation for our case is in terms of
spiral wave solutions and their tips. Suppose the conditions
$\mu_1(\V)=0$, $\mu_2(\V)=0$ determine that the tip of the spiral wave
is located at the origin, and condition $\mu_3(\V)=0$ fixes its
orientation, so $\M$ consists of such functions that look like spiral
waves $\V$ which have the tip exactly at the origin and in a standard
orientation. Then equation \eq{manifold} states that any spiral wave
solution $\u(\r)$, considered at a fixed moment of time, can be
transformed by a Euclidean transformation, in a unique way, to a
solution $\v(\r)$ which has its tip at the origin and in the standard
orientation. This is equivalent to saying that $\v(\r)$ is the same
as $\u(\r)$ only considered in a different system of coordinates, with
the origin at the tip of $\u(\r)$ and oriented accordingly to the
orientation of that tip. We shall say this is the system of
coordinates \emph{attached to the tip}. An example of $\mu_\ell$, as used
\eg\ in \cite{Biktashev-etal-1996}, is
\begin{subequations}
                                                            \eqlabel{mus}
\begin{align}
  \mu_1[\v(\r)]=v^{(l_1)}(\vzero)-\ustar,                   \eqlabel{mu1}\\
  \mu_2[\v(\r)]=v^{(l_2)}(\vzero)-\vstar,                    \eqlabel{mu2}\\
  \mu_3[\v(\r)]=\partial_xv^{(l_3)}(\vzero),                \eqlabel{mu3}
\end{align}
where $\{l_1,l_2,l_3\}\subset\{1,\dots,n\}$ are suitably chosen components, and
$l_1\ne l_2$. This means that the tip of $\u(\r)$ is defined as the point
of intersection of isolines of the components $l_1$ and $l_2$ of the field
$\u$ at appropriately chosen levels $\ustar$ and $\vstar$
respectively, and the orientation of the attached coordinate system is
such that gradient of component $l_3$ (which may or may not coincide
with $l_1$ or $l_2$) is along the $y$-axis in that system. This choice of
$\mu_\ell$ is of course not prescriptive, and later in this paper we
shall consider some variations.

Regardless of the exact definition of the tip, \ie\ choice of
functionals $\mu_\ell$, an essential assumption that we have to make is
that our spiral waves have one tip only, otherwise there would be more
than one way to transform them to the standard position or to chose the
attached system of coordinates. Hence the reason for a further
constraint to the subset $\S\subset\B_0$, which we now can
define as consisting of such one-tip spiral wave solutions, or
functions that look like it:
without such constraint, the whole set $\B_0$ includes
solutions with no tips or more than one tip, for which the
decomposition \eq{manifold} would not hold.
As before, we assume that set $\S$ is
invariant with respect to the dynamic equation \eq{rdeU} for not too
big $||\H||$, that is, if $\U(0)\in\S$, then $\U(t)\in\S$
for all $t>0$ and $||\H||<H_{\max}$.

A further restriction is on the manifold $\M$. It is easy to see
that equations like \eq{mus} may not be sufficient
to define the
manifold with the required property that any orbit crosses it only
once. For instance, if $\v(\r)$ satisfies \eq{mus}, then $\v(-\r)$
also satisfies it, so a rotation by $180^\circ$ around the origin
transfers a point on $\S$ to another point on $\S$.  So to make
the representation \eq{manifold} unique, rather than just requiring
that the gradient of the $l_3$-component of $\v(\r)$ is along the $y$
axis, one would need to specify in which direction it is, say add to
the definition of $\M$ by the equations $\mu_{1,2,3}[\v]=0$
a further inequality 
\begin{equation}
  \mu_4[\v]>0, 
  \quad \textrm{where} \quad
  \mu_4[\v(\r)]=\partial_yv^{(l_3)}.
                                                            \eqlabel{mu4}
\end{equation}
\end{subequations}
This comment extends to the
variations of \eq{mus} which we consider later.

By performing the decomposition \eq{manifold} for every $t\ge0$, we
decompose motion in $\S$ to motion along the RM and motion along
group orbits which are transversal to the RM (see illustration in
\fig{theory-banach}).

\sglfigure{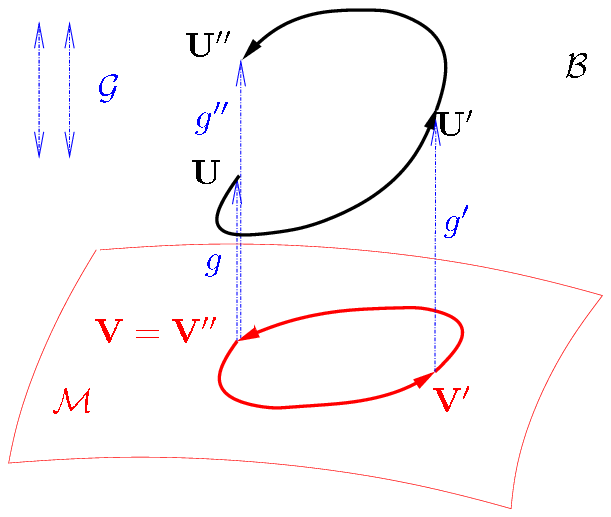}{
  (Color online)
  Sketch of 
  skew-product decomposition of an equivariant flow using a
  Representative Manifold $\M$, which has exactly one transversal
  intersection with evey group orbit $g\in\G$ within the
  relevant stratum of the phase space $\B$ and is diffeomorphic to the
  orbit manifold. Trajectory $(\U,\U',\U'')$ of an equivariant flow in $\B$
  is a relative periodic orbit, since it
  projects onto the trajectory $(\V,\V',\V''=\V)$ on $\M$ which is periodic.
  The flow on $\M$ is devoid
  of symmetry $\G$. 
}{theory-banach}

So for all $t\ge0$, we have 
\begin{equation}
  \U(t) = \T(\g)\V(t)                                       \eqlabel{UTV}
\end{equation}
Substituting \eq{UTV} into \eq{rdeU} and applying $\T\left(\g^{-1}\right)$ to both
sides, we get
\begin{equation}
  \T(\gi)\Df{\T(\g)}{t}\V+\df{\V}{t} = \F(\V)+\Htil(\V,\g,t)      \eqlabel{rdeV}
\end{equation}
where
\begin{equation}
  \Htil(\V,\g,t) = \T(\gi)\H(\T(\g)\V,t)                    \eqlabel{tilH}
\end{equation}
We note that if $\H=\bzero$, the right-hand side of \eq{rdeV} is 
independent of $\g$.

By the assumptions made, intersection of the group orbit $\T(\G)\V$ with
the manifold $\M$ at the point $\V$ is transversal. This means that the
vectors $\F(\V)$ and $\Htil(\V,\g,t)$ can be uniquely decomposed
into the sums of the components along the group and along the
manifold,
\begin{subequations}
  \begin{align}
    \F(\V) &= \Fg(\V)+\Fm(\V),                              \eqlabel{Fsplit}\\
    \Htil(\V,\g,t) &= \Hg(\V,t)+\Hm(\V,t).                  \eqlabel{Hsplit}
  \end{align}
                                                            \eqlabel{FHsplit}
\end{subequations}

Hence equation \eq{rdeV} splits into two components, along the RM and
along the group orbit (GO):
\begin{subequations}
  \begin{align}
    &\textrm{(RM)} && \df{\V}{t} = \Fm(\V)+\Hm(\V,t) \eqlabel{mani} \\
    &\textrm{(GO)} && \T(\gi)\Df{\T(\g)}{t}\V = \Fg(\V)+\Hg(\V,t) \eqlabel{group}
  \end{align}
                                                            \eqlabel{decomposed}
\end{subequations}
Note that equation \eq{mani} is the equation on the
infinite-dimensional manifold $\M$, \ie\ corresponds to a partial
differential equation, whereas the left- and right-hand sides of the
equation \eq{group} are in the tangent space to the
finite-dimensional group orbits, and the dynamic variable $\g$ is an
element of the finite-dimensional manifold $\G$, so \eq{group} is
in fact a system of ordinary differential equations of order $d=\dim\G$.

  At this point we comment on what we see as a significant difference between our
  approach and that proposed by Beyn and
  Thummler~\cite{Beyn-Thummler-2004} (BT).  Using our notation, in
  place of our ``pinning'' conditions $\mu_\ell(\V)=0$, $\ell=1,\dots
  d$, they defined ``phase conditions'' of the form $\mu_\ell(\V,\g)=0$
  (see equation (2.22) in \cite{Beyn-Thummler-2004}), subsequently
  further generalized to
  $\mu_\ell(\V,\g,\d\g/\d{t})=0$ (\ibid, equation (2.33)). 
  This means that their decomposition $\U=\T(\g)\V$ is not uniquely
  determined by the current state $\U$, but depends on history.
  Such
  generalization may have its advantages and, apparently, works well
  for relative equilibria, \ie\ steadily rotating
  spirals~\cite{Beyn-Thummler-2004,Hermann-Gottwald-2010}. However,
  the situation is different if the solution is a meandering spiral,
  \ie\ is periodic with period $P$ in the orbit space (as illustrated
  in \fig{theory-banach}).  This means that $\U(t+P)$ is equivalent to
  $\U(t)$ up to some Euclidean transformation. In our approach, it is
  then guaranteed, that $\V(t+P)=\V(t)$, as by \eq{manifold},
  $\mu_\ell(\T(\gi)\U)=0$ has
  a unique solution for $\g$ at a given $\U$.  However, in the BT approach, typically
  $\V(t+P)\ne\V(t)$, since $\mu_\ell(\T(\gi)\U,\g,\d\g/\d{t})=0$ does
  not uniquely define $\g$, as $\d\g/\d{t}$ is
  not fixed. So in our approach, study of meandering
  spirals reduces to study of periodic solutions for $\V(t)$, but it
  does not do so in the BT approach.

A practical approach to the problem of decomposing the vector fields
as in \eq{FHsplit} is as follows. Equations \eq{decomposed}
together with the definition of the RM via functionals
$\mu_\ell$ can be re-written in an equivalent form
\begin{subequations}
                                                  \eqlabel{man}
  \begin{align}
    &\df{\V}{t} = \F+\Htil+\A,                   \eqlabel{along-man}\\
    &\mu_\ell(\V(t)) = 0, \quad \ell=1,\dots,d,   \eqlabel{on-man} \\
    &\T(\gi)\Df{\T(\g)}{t}\V = -\A,               \eqlabel{across-man}
  \end{align}
\end{subequations}
where $\A=\A(\V,t)=-\Fg(\V)-\Hg(\V,t)$ is a vector 
belonging to the three-dimensional 
tangent space of the group orbit $\T(\G)(\V)$ at $\V$.
In this formulation, at any
given moment of time, equations \eq{along-man} and \eq{on-man}
together define the evolution of $\V$ and the current value of the
vector $\A$, whereas equation \eq{across-man} defines the
evolution of $\g$.

By definition, vector $\A$ is a result of action of 
a linear combination of the generators of the Lie group $\T(\G)$
as linear operators on $\V$. 
To write the explicit expression for the general form $\A$ 
for our case, let us introduce coordinates $(\R,\Theta)$
on $\G=\SEtwo$, where $\R=(X,Y)$ is the translation vector,
$\Theta$ is the rotation angle and a group element acts as
\begin{equation}
  \g = (\R,\Theta) \; : \; \r \mapsto \R + e^{\mxi\Theta} \r ,
\end{equation}
where $\mxi=\Mx{0&-1\\1&0}$, so $\exp(\mxi\Theta)$ is the matrix
of rotation by angle $\Theta$. 

Using this representation, differentiating the definition of $\T(\g)\v$
given by \eq{T}, and substituting the result into \eq{across-man}, 
we get
\begin{equation}
  \A = \omega \partial_\theta\v + (\c\cdot\nabla)\v ,
\end{equation}
where
\begin{equation}
  \omega=\dot\Theta, \qquad \c=e^{-\mxi\Theta}\dot\R ,
                                                            \eqlabel{quotdiff}
\end{equation}
and $\theta$ is the polar angle in the $(x,y)$ plane, so
$\partial_\theta=x\partial_y-y\partial_x$.

With this result, the system \eq{man} in the original PDE notation
states
\begin{subequations}
  \begin{align}
    & \df{\v}{t} =    \D \nabla^2\v + \f(\v) 
    + \h\left(\v,e^{\mxi\Theta}\nabla\v, \R+e^{\mxi\Theta}\r,t\right) 
  \nonumber \\   & \hspace{5em} 
    + (\c\cdot\nabla)\v + \omega\df{\v}{\theta}, 
                                                            \eqlabel{ezfnum-rda}  \\
    & v^{(l_1)}(\vzero,t) = \ustar, \quad
      v^{(l_2)}(\vzero,t) = \vstar,                       \eqlabel{ezfnum-pina} \\
    & \df{v^{(l_3)}(\vzero,t)}{x} = 0, \quad
      \df{v^{(l_3)}(\vzero,t)}{y} > 0,                    \eqlabel{ezfnum-pinb} \\
    & \Df{\Theta}{t} = \omega , \quad
      \Df{\R}{t} = e^{\mxi\Theta} \c .                      \eqlabel{ezfnum-tip}
 \end{align}                                                \eqlabel{ezfnum}
\end{subequations}
where the dynamic variables are $\v(\r,t)$, $\c(t)$, $\omega(t)$,
$\R(t)$ and $\Theta(t)$.

In terms of the tip of the wave, equation \eq{ezfnum-rda} is the
original reaction-diffusion equation \eq{rdeu} written in the 
comoving FoR,
equations \eq{ezfnum-pina}, \eq{ezfnum-pinb} define the
attachment (pinning)
of the tip to this FoR,
and equations \eq{ezfnum-tip} describe the movement 
of the FoR and, therefore,
of the tip.

  Equations \eq{ezfnum-pina},~\eq{ezfnum-pinb} imply that the position
  $(\xtip,\ytip)$ and orientation $\Phi$ of the tip during calculations
  in the laboratory FoR are defined as
  \begin{subequations}
    \begin{align}
      & u^{(l_1)}(\xtip(t),\ytip(t),t)=\ustar, \\
      & u^{(l_2)}(\xtip(t),\ytip(t),t)=\vstar, \\
      & \Phi(t)=\arg\left( (\partial_x+i\partial_y)u^{(l_3)}(\xtip(t),\ytip(t),t)\right)
    \end{align}
  \end{subequations}
  and the comoving FoR is chosen so that in it, $(\xtip,\ytip)=(0,0)$
  and $\Phi=\pi/2$ at all times. 
Unlike other
equations of system~\eq{ezfnum}, these are not prescriptive and is essentially an
arbitrary choice, dictated by properties of particular systems. 
We shall refer to the pinning
  conditions~(\ref{eq:ezfnum-pina},\ref{eq:ezfnum-pinb}) as ``Choice 1'', as below
  we shall consider a variation of these, which we call
  ``Choice 2''. 

When $\h=\bzero$, the system \eq{ezfnum} decouples, as its upper part
including \eq{ezfnum-rda},  \eq{ezfnum-pina} and \eq{ezfnum-pinb} becomes
independent of the lower part \eq{ezfnum-tip}. This is the
``skew-product'' decomposition, the upper part describing the dynamics
in the space of group orbits, so called ``quotient system'', and the lower
part the ``symmetry group extension'', \ie\ 
dynamics along the group, which depends on but does not
affect the quotient dynamic. The connection between the quotient system
and the group extension is via the dynamic variables $(\c,\omega)$;
in the following, we refer to these three quantities as 
``quotient data'' for brevity.  
 
The skew-product representation has been useful for
the analysis of various types of meander of spiral waves
\cite{Biktashev-etal-1996,Wulff-1996,Biktashev-Holden-1998,Nicol-etal-2001}. 
Note that the approach used in
\cite{Wulff-1996,Nicol-etal-2001} (also see references therein) is based on the
assumption that the meandering pattern in question is considered in
the vicinity of a bifurcation from the rigidly rotating spiral wave
solution, so that the quotient dynamics can be reduced to the centre
manifold, hence instead of equations \eq{ezfnum-rda},
\eq{ezfnum-pina} and \eq{ezfnum-pinb}, these studies considered
normal forms on the corresponding centre manifolds. However, as noted
in \cite{Ashwin-etal-2001}, the Centre Manfold Theorem is not applicable for
spiral waves, so this approach seems to be fundamentally flawed. This
technical difficulty of course does not in any way affect the validity
of system \eq{ezfnum}, which, as we have just demonstrated, is derived
by elementary means without recourse to any bifurcations. 

In the rest of the paper, we consider system \eq{ezfnum} as a
computational tool, rather than an instrument of theoretical analysis.
The disadvantage of the original system \eq{rdeu} as a computational
tool is that it requires a big computational grid to simulate dynamics
of a spiral in an infinite medium, particularly when the tip of the
spiral performs excursions to large distances. This is actually not
necessary, as the dynamics of the spiral is mostly determined by the
events in some finite vicinity of its tip \cite{Biktasheva-Biktashev-2003}. The system
\eq{ezfnum} takes advantage of this property so that the PDE
calculations are done always in some fixed vicinity of the spiral
wave, whereas the movement of the tip is described by the ODE part.


\section{Numerical Implementation}\seclabel{numerics}

\paragraph{Discretization.} We use time discretization with constant
step $\stept$ and square spatial grid with step $\stepx$, covering
spatial domain $(x,y)\in[-\L/2,\L/2]^2$, so that
\[ 
  \v((i-i_0)\stepx,(j-j_0)\stepx,k\stept)
  \sim\vhat^{k}_{i,j}
\]\[
  =\left(\hat v^{(l),k}_{i,j} \,|\, l=1,\dots,n\right), \quad 
  i=0,\dots\Nx, \; j=0,\dots\Ny, 
\]\[
  \Nx=\Ny=\L/\stepx ,
\] 
and the grid
coordinates of the origin are 
\[
  i_0=(\Nx+1)/2, \; j_0=(\Ny+1)/2
\]
(we only use odd values of $\Nx=\Ny$). 
We designate
the $k$-th time layer of the numerical solution as
$\Vhat^k=\left(\vhat^{k}_{i,j} \,|\, i=1,\dots\Nx,j=1,\dots\Ny\right)$. We
discretize the ODE dynamic variables on the same time grid, \ie\
$\R(k\stept)\sim\Rhat^k$ etc.

\paragraph{Operator splitting.}

We rewrite 
equation \eq{ezfnum-rda} in the form
\begin{equation}
   \df{\v}{t} = \Fcal[\v] + \Hcal[\v;\R,\Theta]  + \Acal[\v;\c,\omega]
\end{equation}
where differential operators $\Fcal$, $\Hcal$ and $\Acal$ are defined as
\begin{subequations}
\begin{align}
   & \Fcal[\v] =  \D \nabla^2\v + \f(\v) ,                  \eqlabel{RDpart} \\
   & \Hcal[\v;\R,\Theta] = \h(\v,e^{\mxi\Theta}\cdot\nabla\v, \R+e^{\mxi\Theta}\r,t), 
                                                            \eqlabel{Ppart} \\
   & \Acal[\v;\c,\omega] = (\c\cdot\nabla)\v + \omega\df{\v}{\theta} \nonumber \\
  & \phantom{\Acal(\v;\c,\omega)} =
   (c_x-\omega y)\,\df{\v}{x} + (c_y+\omega x)\,\df{\v}{y} .
                                                            \eqlabel{Apart}
\end{align}
\end{subequations}
Let $\Fhat$, $\Hhat$ and $\Ahat$ be discretizations of $\Fcal$, $\Hcal$ 
and $\Acal$. Our computations proceed as follows:
\begin{subequations}
\begin{align}
  & \Vhat^{k+1/3} = \Vhat^{k} + \stept\Fhat\left(\Vhat^{k}\right) , 
                                                            \eqlabel{RDstep} \\
  & \Vhat^{k+2/3} = \Vhat^{k+1/3} + \stept\Hhat\left(\Vhat^{k+1/3},\Rhat^{k},\Thetahat^{k}\right) ,
                                                            \eqlabel{Pstep} \\
  & \Vhat^{k+1}   = \Vhat^{k+2/3} + \stept\Ahat\left(\Vhat^{k+2/3},\chat^{\;k+1},\omegahat^{k+1}\right) ,
                                                            \eqlabel{Astep} \\
  & \mu_{1,2,3}\left(\Vhat^{k+1}\right) = 0 , \; \mu_4\left(\Vhat^{k+1}\right)>0, \eqlabel{mustep} \\
  & \Thetahat^{k+1} = \Thetahat^k + \stept \, \omega^{k+1},             \eqlabel{Thetastep} \\
  & \Rhat^{k+1} = \Rhat^k + \stept \, e^{\mxi\Thetahat^{k+1}}\c\,^{k+1} \eqlabel{Rstep} .
\end{align}                                                 \eqlabel{step}
\end{subequations}

\paragraph{Kinetics.}
As specific examples, we consider two models, the FitzHugh-Nagumo
model \cite{FitzHugh-1961,Nagumo-etal-1962}:
\begin{equation}
  \f: \Mx{u\\v} \mapsto \Mx{ 
    \alpha^{-1}\left(u-u^3/3-v\right) \\ 
    \alpha(u+\beta-\gamma v)
  }
\end{equation}
and Barkley's model \cite{Barkley-etal-1990,Barkley-1991}:
\begin{equation}
  \f: \Mx{u\\v} \mapsto \Mx{
    c^{-1}u(1-u)\left(u-(v+b)/a\right) \\
    u - v
  }
\end{equation}
both with $\D=\Mx{1&0\\0&0}$. 

\paragraph{Reaction-diffusion step.}

The computational scheme is designed as an extention to the standard
approach to simulation of spiral waves. Specifically, we chose
Barkley's \ezspiral\  \cite{Barkley-etal-1990,Barkley-1991,EZSpiral} as the starting point, and
extended it to add the other computational steps. So the
reaction-diffusion step \eq{RDstep} is as implemented in \ezspiral,
with central 5-point difference approximation of the Laplacian, 
without any features specific to the Barkley model, such as
  implicit treatment of the kinetic terms,
and
with appropriate modifications when FitzHugh-Nagumo model is used.

\paragraph{Perturbations.}

We consider one particular type of nonzero perturbation, the
electrophoresis, 
\begin{align}
  \h &=\E\partial_x\u, \nonumber\\
 \htil &=\E\left(\cos(\Theta)\partial_x\v(r)-\sin(\Theta)\partial_y\v(r)\right),
                                                            \eqlabel{perturbation}
\end{align}
where $\E$ is a diagonal matrix, $\E=\Mx{E_1&0\\0&E_2}$, $||\E||\ll1$. 
For a reaction-diffusion system this perturbation can describe
movement of the reagents in response to electric field with velocities
$-E_1$ and $-E_2$ along the $x$-axis. 
For $\E=\epsilon\D$, this perturbation can also approximately describe the movement
of an axially symmetric scroll ring. For a cylindrical system of
coordinates $(r,\theta,z)$: $x=r\cos\theta$, $y=r\sin\theta$, $z=z$,
the diffusion term has the form
$\D\nabla^2\u=\D(\partial_r^2+\frac{1}{r}\partial_r+\frac{1}{r^2}\partial_\theta^2+\partial_z^2)\u$,
which for $\partial_\theta=0$ and large $r$ is equivalent to an
unperturbed diffusion term with a 2-dimensional Laplacian in $(r,z)$
plane plus a small perturbation $\frac1r \D \partial_r \u$. If the
filament of the scroll is located at large values of
$r\approx1/\epsilon$ and as the dynamics of the scroll is mostly
determined by the events near its filament, then $1/r$ can be
approximately replaced with $\epsilon$.

Perturbation \eq{perturbation} violates
only rotational symmetry of the problem, preserving symmetry with
respect to translations in space and time. Hence $\htil$ explicitly
depends only on $\Theta$. This limitation is not principal and translation
symmetry breaking perturbations can be considered similarly, in which
case $\htil$ would also explicitly depend on $X,Y$ and/or $t$. We
discretize the first spatial derivatives in the perturbation term
using upwind second-order accurate differences, and use explicit Euler
timestepping. In the absence of perturbations, $\h=\bzero$, the
perturbation step \eq{Pstep} is of course omitted and
$\Vhat^{k+2/3}=\Vhat^{k+1/3}$.

\paragraph{Tip definition and pinning conditions.}
Discretization of the pinning conditions~\eq{ezfnum-pina},~\eq{ezfnum-pinb}, 
using $l_1=l_3$, and the right-side first-order discretization 
of the $x$-derivative, gives
\begin{subequations}
                         \eqlabel{musteps}
\begin{align}
  \hat v^{(l_1),k}_{i_0,j_0}=u_* , \eqlabel{mustep1} \\
  \hat v^{(l_2),k}_{i_0,j_0}=v_* , \eqlabel{mustep2} \\
  \hat v^{(l_1),k}_{i_0+1,j_0}=u_* , \eqlabel{mustep3}
\end{align}
\end{subequations}
where $(i_0,j_0)$ are grid coordinates of the origin. This works in
principle, but gives rather inaccurate and noisy approximations for
$\omega$, which get worse for finer discretizations. This is typical
for numerical differentiation. We overcome this by enhancing the
spatial discretization step, by replacing the condition \eq{mustep3}
with
\begin{equation}
 \hat v^{(l_1),k}_{i_1,j_1}=u_* , \eqlabel{mustep3a}
\end{equation}
where the grid point $(i_1,j_1)$ was chosen some way away from the
centre point, $(i_1,j_1)=(l_0,j_0)+(\iinc,\jinc)$. 
This means replacing the third pinning
  condition~\eq{ezfnum-pinb} with
   \begin{equation}
      v^{(l_1)}(\rinc,t) = \ustar,                    \eqlabel{ezfnum-pinc}
    \end{equation}
 where $\rinc=(\stepx\iinc,\stepx\jinc)$. 
Empirically, we have found that the length of the displacement
$|\rinc|$ should be of the order of, but not
exceeding, one full wavelength of the spiral. 

This revised
orientation-pinning condition still does not define the position
uniquely, as illustrated by \fig{ezfnum-unique1}. An extra inequality
is required to distinguish between 
different solutions satisfying conditions
(\ref{eq:mustep1},\ref{eq:mustep2},\ref{eq:mustep3a}).
We use
\begin{equation}
 \hat v^{(l_1),k}_{i_1,j_1} < v_* . \eqlabel{mustep4}
\end{equation}
corresponding to
\begin{equation}
 v^{(l_1),k}(\rinc,t) < v_* . \eqlabel{ezfnum-pind}
\end{equation}
Specifically, we chose $l_1=l_3=1$ and $l_2=2$. Conditions
\eq{mustep4} and \eq{ezfnum-pind}
then mean that the third pinning condition (\ref{eq:mustep3a},\ref{eq:ezfnum-pinc}) ensures that
the front, rather than the back, of the excitation wave passes though
the grid point $(i_1,j_1)$.
So
  equations~(\ref{eq:ezfnum-pina},\ref{eq:ezfnum-pinc},\ref{eq:ezfnum-pind}),
  with their discretizaions (\ref{eq:musteps},\ref{eq:mustep4}) are our ``Choice
  2'' pinning conditions. 

The Choice 1 and Choice 2 pinning conditions define
  different RMs and different quotient data $\c(t),\omega(t)$, for the
  same solution $\u(\r,t)$. However, the two FoRs they define have a
  common origin and differ only by the orientation angle. So if
  $(\c,\omega)$ are quotient data for Choice 1 pinning conditions, and
  $(\c\,',\omega')$ are quotient data for Choice 2 pinning conditions,
  then we have 
  \begin{equation}
    \c=e^{\mxi(\Phi-\pi/2)}\c\,', \qquad \omega=\omega'+\d\Phi/\d{t}, \eqlabel{gauge}
  \end{equation}
  where $\Phi$ the tip orientation angle in the Choice 2 comoving FoR,
  so $\Phi-\pi/2$ is angle of one FoR against the other.

\sglfigure{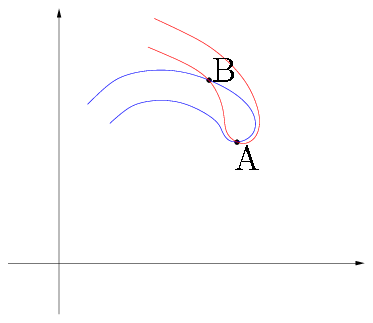}{
  (Color online)
  Non-uniqueness of the revised tip pinning condition.
}{ezfnum-unique1}

\paragraph{Advection.}
We use an upwind second-order accurate approximation of the
spatial derivatives in $\Ahat$. The steps \eq{Astep}
and \eq{mustep} are done in conjunction with each other. The discretization of $\Vhat^{k+1}$ at
the tip pinning points, resulting from \eq{Astep}, is used in the
three equations \eq{mustep} to find the three unknowns $\chat_x^{\,k+1}$,
$\chat_y^{\,k+1}$ and $\omegahat^{k+1}$, so that the pinning conditions \eq{mustep}
are always satisfied exactly (to the processor precision) after every
step~\endnote{
  We are thankful to B.N. Vasiev for this idea. 
}.

\paragraph{Boundary conditions.}
Since the boundaries in the comoving FoR do not represent any
physical reality but are only a necessity of numerical approximation,
the results can only be considered to be reliable if they do not depend on
the boundary conditions. So we use both Dirichlet and Neumann bondary
conditions and compare the results. For Dirichlet conditions, we use
boundary values of the resting state $\v_r$, such that
$\f(\v_r)=\bzero$.

\paragraph{Tip trajectory reconstruction.}
The steps \eq{Thetastep} and \eq{Rstep} are simple first-order
implementations of the corresponding ODEs. The resulting $\Thetahat$
is used in calculations of the $\Hcal$ step when the perturbation is on.
Otherwise, $\Thetahat$ and $\Rhat$ are calculated only for the record.

\paragraph{Some details of software implementation.} 
For stability purposes, we ensure that the following inequalites are
observed during computations, 
\begin{eqnarray*}
|c_x| &\leq& \frac{\stepx^2}{2\stept} , \\
|c_y| &\leq& \frac{\stepx^2}{2\stept} , \\
|\omega| &\leq& \frac{1}{N_X\stept} .
\end{eqnarray*}
This is an empirical choice motivated by von Neumann stability
analysis.

When the absolute values of $c_x$ and $c_y$ found in \eq{Astep} and
\eq{mustep} are beyond these limits then they are
restricted to the intervals stated above. Also, we eliminated the need
to restrict the values of $c_x$ and $c_y$ to their stability limits by
moving the spiral wave solution so that the tip of the spiral wave is
in the center of the box, using the standard \ezspiral's `mover'
function, which performs translation of the solution by an integer
number of grid steps, suitably extrapolating the solution where
necessary near the boundaries.

For $\omega$, we implemented the restriction that if $|\omega|$
exceeded its maximum stability value, then then we set $\omega=0$. 
Effectively this means that unless the orientation of the spiral wave
is already very near the standard orientation satisfying equation
\eq{mustep3a} and inequality \eq{mustep4}, the code computes
a solution of the problem
\begin{subequations}
  \begin{align}
    & \df{\v}{t} =    \D \nabla^2\v + \f(\v) 
   + \h\left(\v,\nabla\v, \R+e^{\mxi\Theta}\r,t\right) 
  \nonumber \\   & \hspace{5em} 
    + (\c\cdot\nabla)\v, 
                                                            \eqlabel{ezfnumt-rda}  \\
    & v^{(l_1)}(\vzero,t) = \ustar, \quad
      v^{(l_2)}(\vzero,t) = \vstar,                         \eqlabel{ezfnumt-pin} \\
    & \Df{\R}{t} = e^{\mxi\Theta} \c                        \eqlabel{ezfnumt-tip}
 \end{align}                                                \eqlabel{ezfnumt}
\end{subequations}
instead of \eq{ezfnum}. That is, it performs reduction by the subgroup
of translations of the Euclidean group. 

A typical run of the program in the interactive mode starts from
obtaining a spiral wave solution in the standard ``ride-off'' mode, by
solving initial-value problem \eq{rdeu}. When the spiral wave is
initiated so there is one tip in the solution, the user switches the
program to the ``ride-on'' mode, with calculations according to the
above scheme. On the switch, the program first of all moves the tip of
the spiral to the centre of the box via \ezspiral's  `mover' function,
\ie\ by parallel translations of the solution, supplementing the
missing pieces near boundaries by duplicating the existing boundary
values. From then on, the spiral continues to rotate with its tip
fixed at the centre of the box, thus solving the problem \eq{ezfnumt}.
In this regime, only the first two pinning conditions are satisfied, and
only $c_x$ and $c_y$ are calculated and used, where as $\omega$ is
calculated but replaced with zero, until it falls within the stability
limit and the fourth inequality-type pinning condition is satisfied.
From that point, the program proceeds in the fully engaged mode,
calculating the problem \eq{ezfnum}.


\section{Primary Examples: Rigidly Rotation and Meander}\seclabel{examples}

First we illustrate how our approach works using two examples. One example
uses Barkley model with rigidly rotating spiral waves, and the other
is FitzHugh-Nagumo model with meandering spiral waves.

\sglfigure{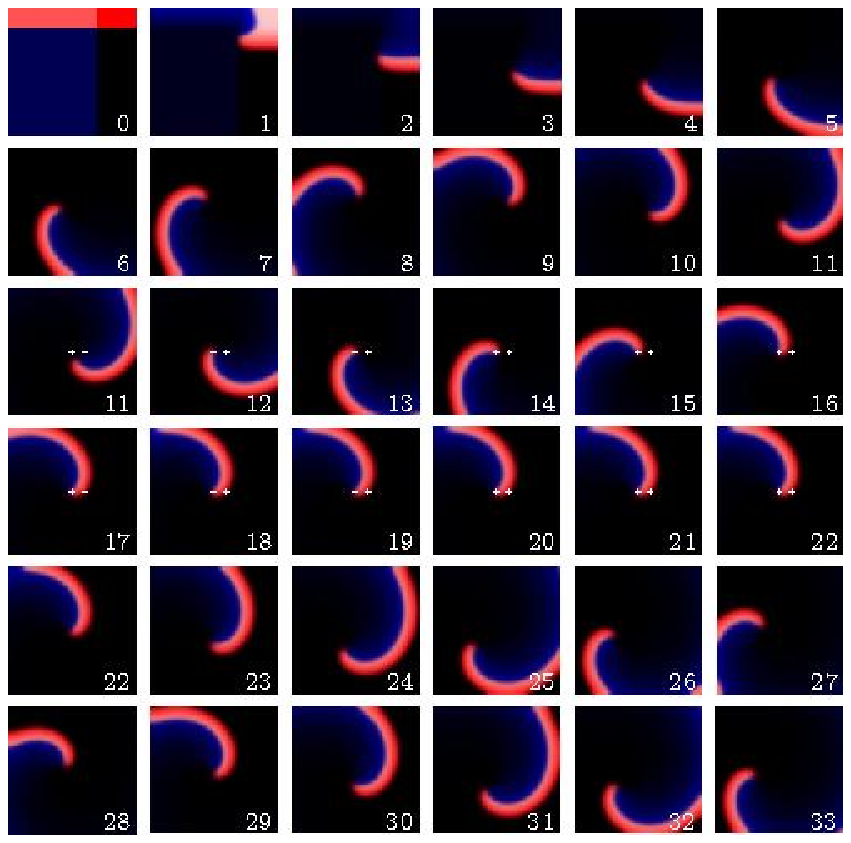}{
  (Color online)
  Three consecutive runs of Barkley model, 
  $a=0.52$, $b=0.05$, $c=0.02$, 
  $\L=20$, $\stepx=1/5$, $\stept=1/2000$, 
  $\rinc=(2,0)$.
  The runs 
  $t\in[0,11]$ and $t\in[22,33]$ are direct simulations.
  The run $t\in[11,22]$ is a quotient system simulation, the pinning points 
  are indicated by small white crosses.
  The third pinning condition is engaged at $t\approx16.5$. 
}{bkl-movie}

\Fig{bkl-movie} illustrates the work of \ezride\  in the case of a
rigidly rotating spiral wave. The panels represent three consecutive
runs, in different regimes: the ``direct numerical simulations'' (DNS)
of system~\eq{rdeu}, then the ``skew-product''
calculation in the comoving FoR, and then again the DNS
in the laboratory FoR. The skew-product calculation in
turn consists of two parts. The first part is described by
\eq{ezfnumt} where only the two translation pinning
conditions are engaged, so that
the position of the tip of the spiral is fixed, but not its
orientation, so the FoR is co-translating but not co-rotating. The
second part is where all four pinning conditions are engaged, and the
FoR is co-translating and co-rotating. It is seen from
\fig{bkl-movie}, that after a transient period, the solution in the
fully comoving FoR becomes stationary. This corresponds to the
definition of a rigidly rotating spiral wave as a relative
equilibrium.

\sglfigure{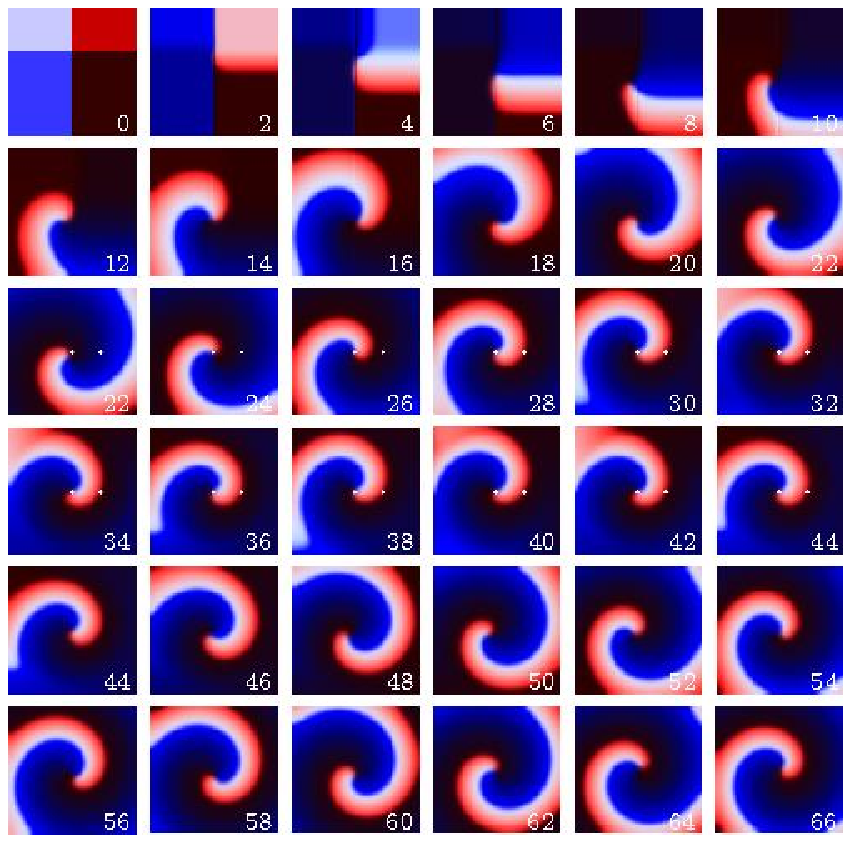}{
  (Color online)
  Three consecutive runs of FHN model, 
  $\alpha=0.2$, $\beta=0.7$, $\gamma=0.5$, 
  $\L=30$, $\stepx=1/3$, $\stept=1/720$,
  $\rinc=(20/3,0)$. 
  The runs 
  $t\in[0,22]$ and $t\in[44,66]$ are direct simulations.
  The run $t\in[22,44]$ is a quotient system simulation, the pinning points 
  are indicated by small white crosses.
  The third pinning condition is engaged at $t\approx27.5$. 
}{fhn-movie}

\Fig{fhn-movie} shows a similar set of runs for a different case,
where the spiral wave is not stationary but is meandering. 
In this case, the solution in the comoving FoR is not stationary,
but periodic in time. This corresponds to the definition of a
meandering spiral wave as a relative periodic orbit.

\sglfigure{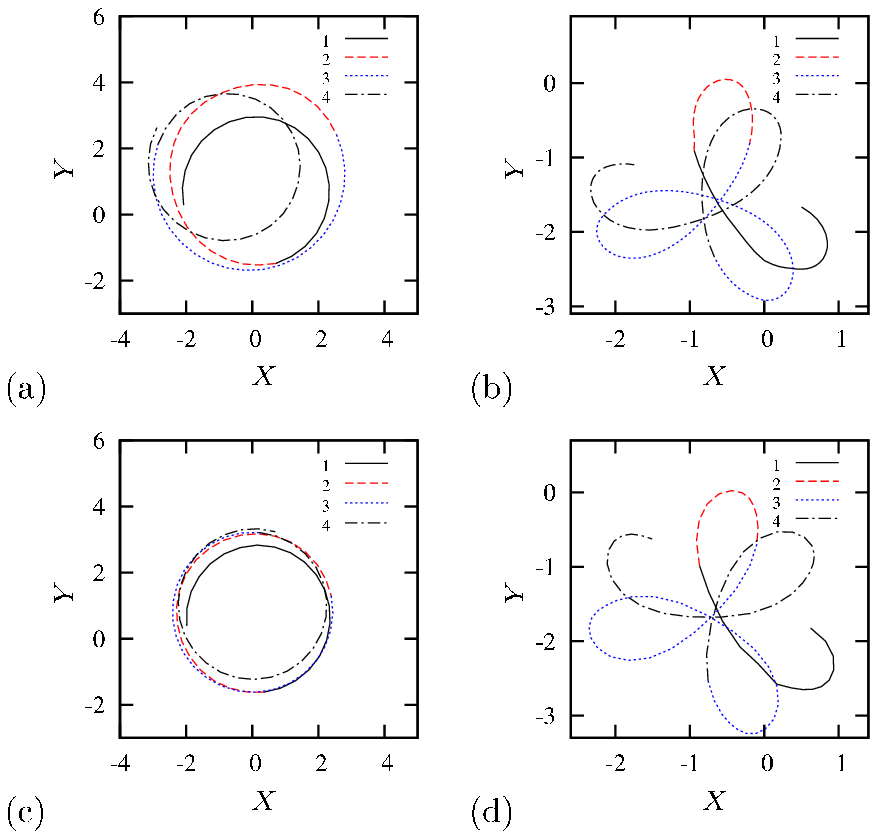}{
  (Color online)
  (a,b)
  Reconstructed tip trajectories from (a) simulation shown in \fig{bkl-movie}
  and (b) simulation shown in \fig{fhn-movie}. The pieces labelled 1 are 
  trajectories obtained
  in direct simulations in the laboratory FoR. 
  The pieces labelled 2 are trajectories obtained via co-translating
  simulations, with first two pinning conditions engaged. 
  The pieces labelled 3 correspond to co-moving (co-translating and co-rotating)
  simulations with all three pinning conditions engaged. 
  The final pieces labelled 4 correspond to direct simulations in a non-moving 
  FoR, which has been displaced with respect to the laboratory FoR during 
  the quotient system simulations.
  (c) Same as (a), with $\stepx=1/10$, $\stept=1/4000$.
  (d) Same as (b), with $\stepx=1/10$, $\stept=1/4000$.
}{clips}

 \Fig{clips}(a,b) show selected pieces of tip trajectories
  obtained as a result of the runs shown in \fig{bkl-movie} and
  \fig{fhn-movie}.  The discretization steps there are
  deliberately chosen crude, to allow very fast running simulations,
  and also to illustrate the difference introduced by the change of
  method of computation. The tip trajectories obtained by
  reconstruction from the quotient data are qualitatively similar
  to the tip trajectories obtained in DNS. However, the quantitative
  difference is also quite evident. In the case of rigid rotation, the
  reconstructed trajectory radius is noticeably bigger than that from
  DNS, and the centres of the meandering patterns in different runs
  are offset against each other.  As panels (c,d) in the same
  figure show, these discrepancies decrease when the discretization
  steps are refined.  

\sglfigure{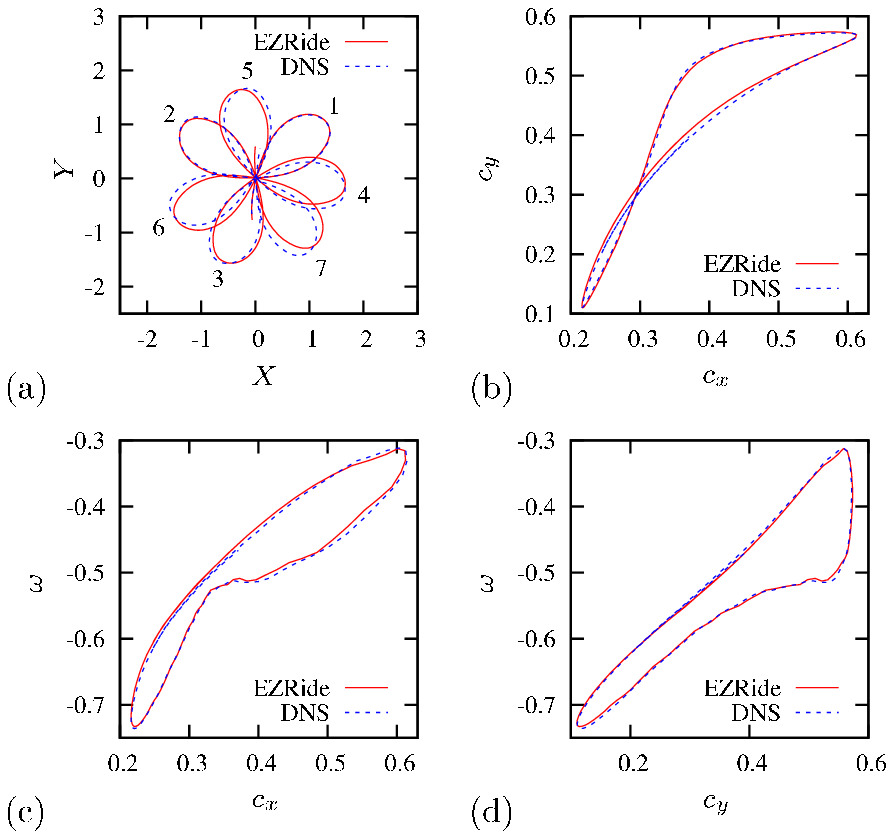}{
  (Color online)
   Meander in the FHN model, calculated in the laboratory
    frame of reference (DNS), and from quotient system (\ezride).  In
    (a), the meandering pattern is shown, which for the \ezride\ curve
    is obtained by numerical integration of quotient data using
    \eq{quotdiff}.  In (b--d), the projections of the quotient data
    are shown, which for the DNS curves are obtained by numerical
    differentiation of the tip trajectory, using \eq{quotdiff}.
}{mean}

 \Fig{mean} shows the tip and quotient system trajectories,
  obtained in laboratory and comoving FoR calculations, for a
  meandering spiral. This is drawn for the finer discretization steps,
  as in \fig{clips}(d). For comparison, quotient data for both the
  laboratory and comoving FoR calculations were recalculated for the
  Choice 1 pinning conditions using \eq{gauge}. 
  There is good agreement between the two
  methods of calculations, within the expected accuracy. More detailed
  analysis of the numerical accuracy of our method is given in the
  next session.

\section{Numerical convergence}\seclabel{convergence}

\sglfigure{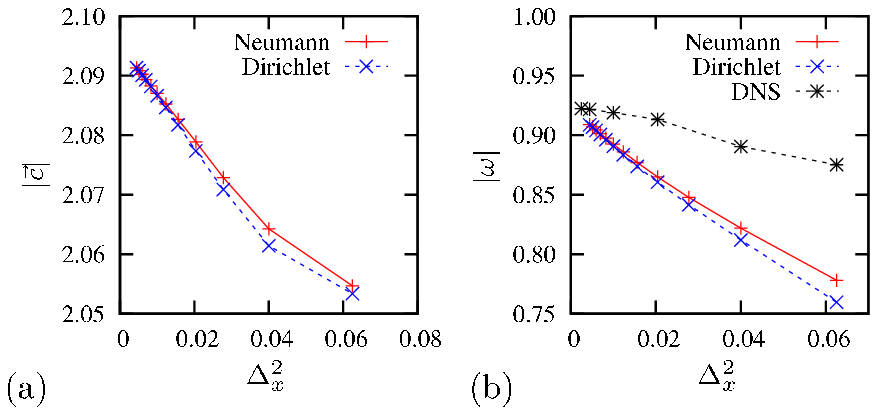}{
  (Color online)
  Convergence of the rigidly rotating spiral wave solution in the
  Barkley's model. 
}{convergence}

\Fig{convergence} illustrates the convergence of the results of
calculations of rigidly rotating spiral, using \ezride\  with Neumann and
Dirichlet boundary conditions, and DNS using Neumann boundary
conditions. In these calculations, the box size is fixed at $\L=60$
and the timestep is changed with the spacestep so that
$\stept=\stepx^2/40$. For $\omega(\stepx^2)$ dependence, we also show
the angular velocity measured in direct numerical simulations. We do
not show $|\c(\stepx^2)|$ found in DNS, since obtaining it involves
numerical differentiation which gives accuracy insufficient
for the convergence study.

Our discretizations are second order accurate in $\stepx$ and first
order accurate in $\stept$ both in DNS and in the riding mode, which
corresponds to linear dependence of any results on $\stepx^2$ for
$\stepx\to0$. We see in \fig{convergence} that this is indeed the
case. Linear extrapolation of the $\omega(\stepx^2)$ gives the values
of $\omega(0)$ for laboratory and comoving calculations coinciding to
within $10^{-3}$.

One of the advantages of \ezride\  is the fact that the simulations can
be done in a smaller box compared to DNS. So, the last test is
convergence in box size. We have calculated the rigidly rotating spiral 
by \ezride\  at fixed $\stepx=1/15$, $\stept=1/9000$ and $\L$ varying through $[15,60]$
and found that both $|\c|$ and $|\omega|$ vary by less than $10^{-3}$. 


\section{Application I: The 1:1 Resonance in Meandering Spiral Waves}\seclabel{resonance}

One of the cases where the DNS would meet with difficulties, is the
study of the the meandering of spiral waves for parameters near the
``1:1 resonance'' between the Euclidean and the Hopf frequencies.
This case is marginal between meandering patterns 
with inward petals and outward petals. Near the resonance,
the spatial extent of the meandering trajectory becomes large, and
for the case of exact resonance, infinite, and the spiral appears to
be spontaneously drifting~\cite{Zykov-1986,Barkley-1994}.
Hence, following the dynamics of the spiral 
wave in the comoving FoR presents an advantage. 

We illustrate this using the FHN model. We fix the discretization
parameters at $\stepx=1/8$, $\stept=1/2560$ and $\L=20$. The choice of
model parameter is influenced by Winfree's ``Flower Garden''
\cite{Winfree-1991}, which gives a rough estimate for the location of the
1:1 resonance line in the $(\alpha,\beta)$ plane at $\gamma=0.5$.
Using this information, we have selected two values $\alpha=0.2$ and
$\alpha=0.25$, and scanned values of $\beta$ across the resonance
value, which we determined as $\beta_0\approx 0.93535$ for
$\alpha=0.2$, and $\beta_0\approx 0.81362$ for $\alpha=0.25$ at our
discretization parameters.

\sglfigure{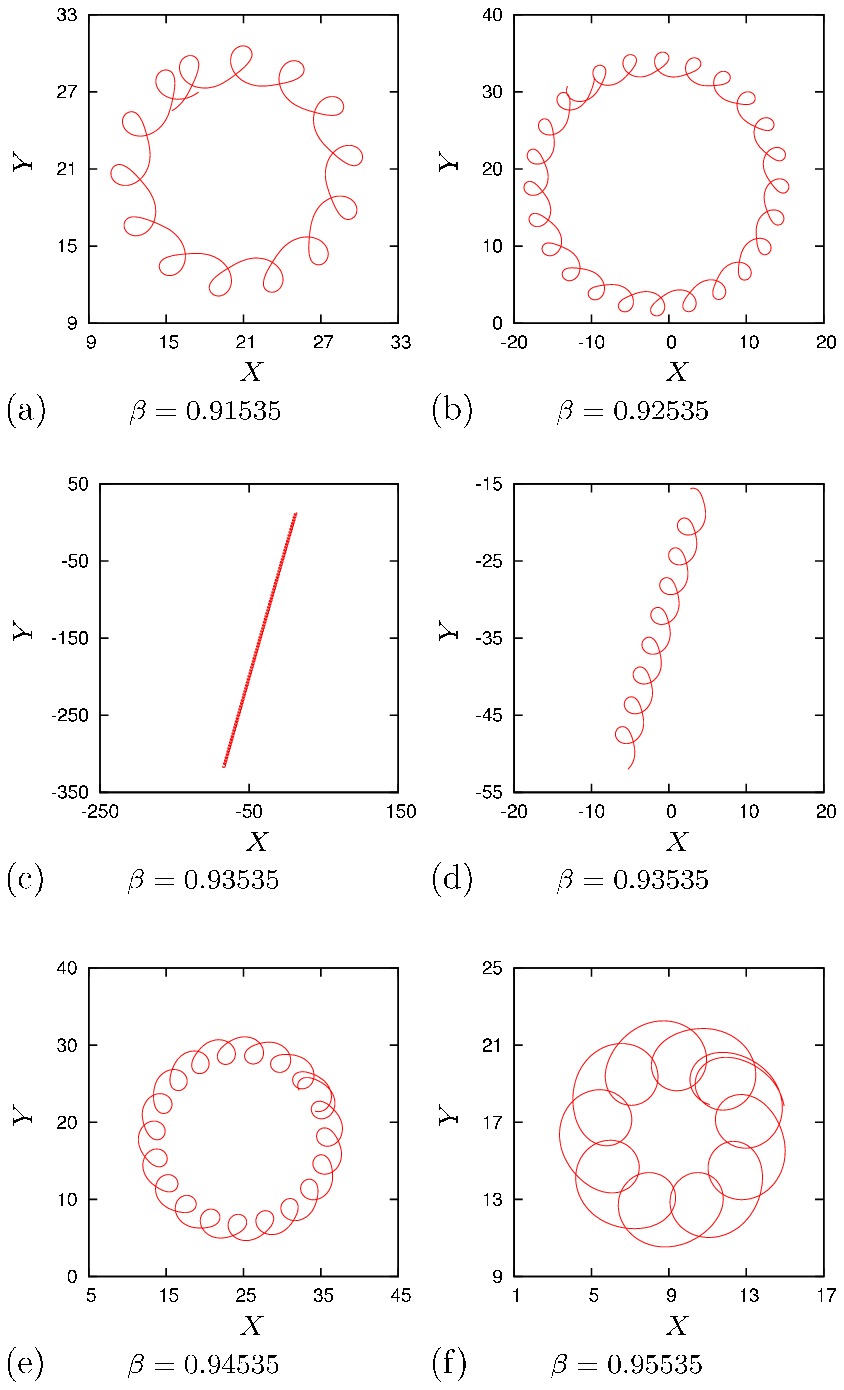}{
  (Color online)
  The reconstructed tip trajectories in FitzHugh-Nagumo 
  system with $\alpha=0.2$, $\gamma=0.5$ and varying $\beta$. 
}{eps02-tip}

\dblfigure{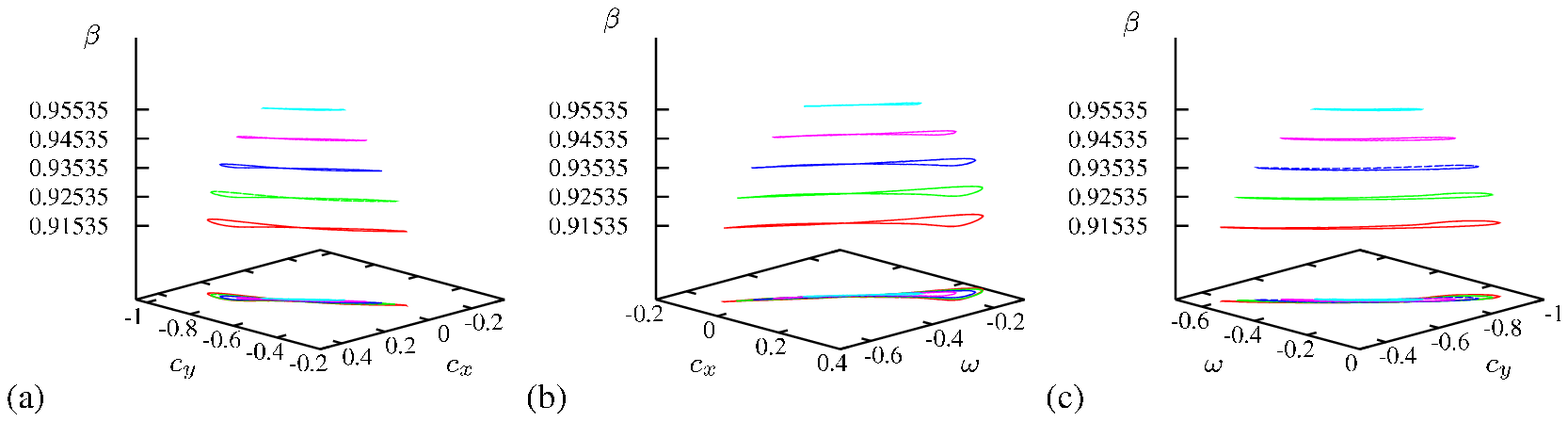}{
  (Color online)
  Various projection of the limit cycles in the quotient system
  corresponding to the trajectories shown in \fig{eps02-tip}. 
}{eps02-quot}

The results are presented on \figs{eps02-tip}--\figref{eps025-quot}.
The shape of trajectories is well known from the theory, and is
outward petals for $\beta<\beta_0$ and inward petals for
$\beta>\beta_0$, degenerating into spontaneous straightforward drift
at $\beta=\beta_0$. The trajectory at $\beta=\beta_0$ in
\fig{eps02-tip} is shown twice: once for the whole duration as it was
calculated, \fig{eps02-tip}(c), and then a close-up of small part of
it, \fig{eps02-tip}(d). Calculation of this particular trajectory
using DNS would require, by our estimate, about five weeks, as opposed
to 2.5~hours used by \ezride.

The change of the quotient dynamics with changing $\beta$
is illustrated in \fig{eps02-quot}. As opposed to the tip
trajectories, there is no evident qualitative changes in the shape of
the limit cycle across $\beta=\beta_0$. Note the very elongated shape
of the limit cycles in all three projections.  We do not
know whether this has some theoretical explanation or is merely
incidental.

\sglfigure{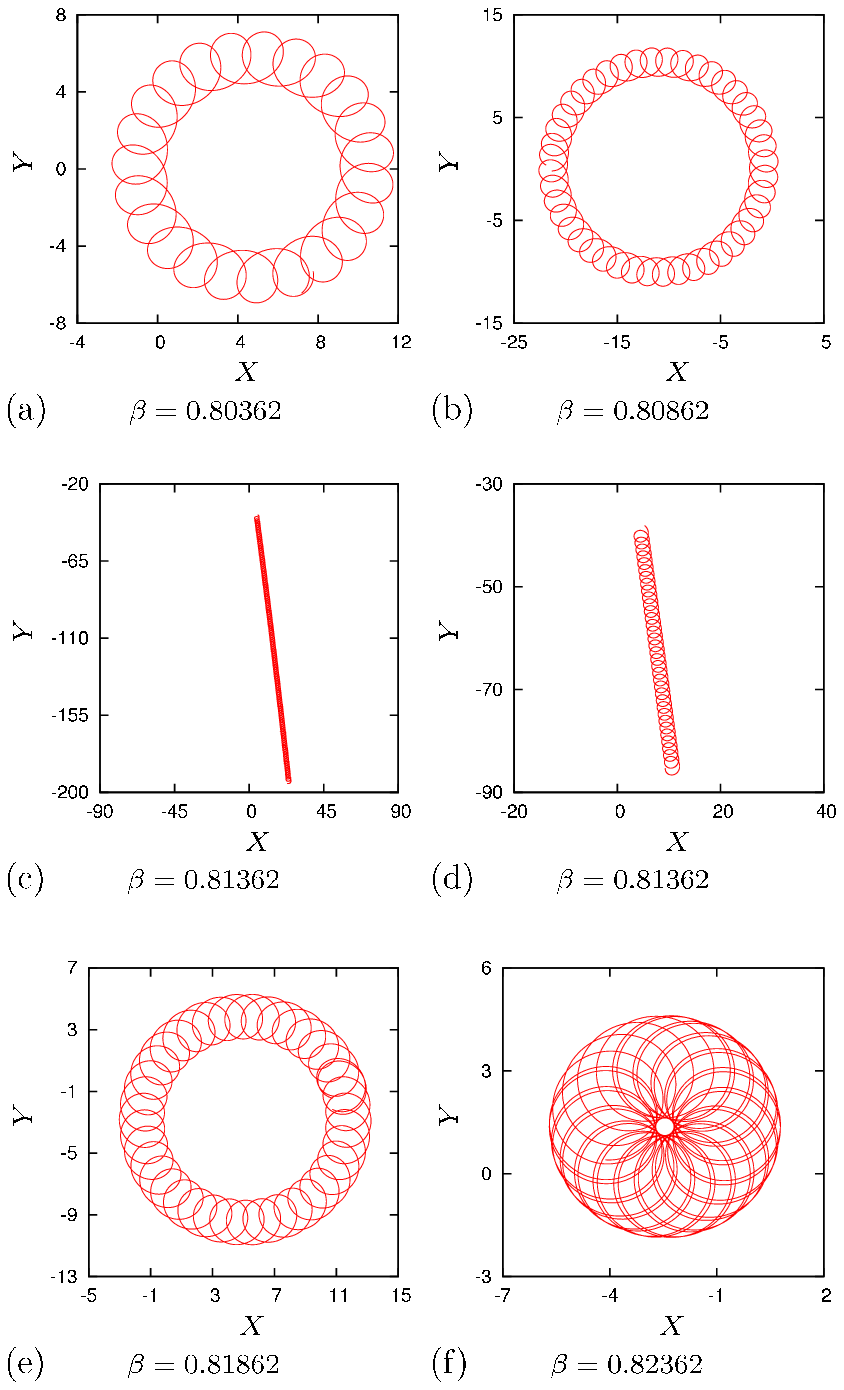}{
  (Color online)
  Same as \fig{eps02-tip}, for $\alpha=0.25$. 
}{eps025-tip}

\dblfigure{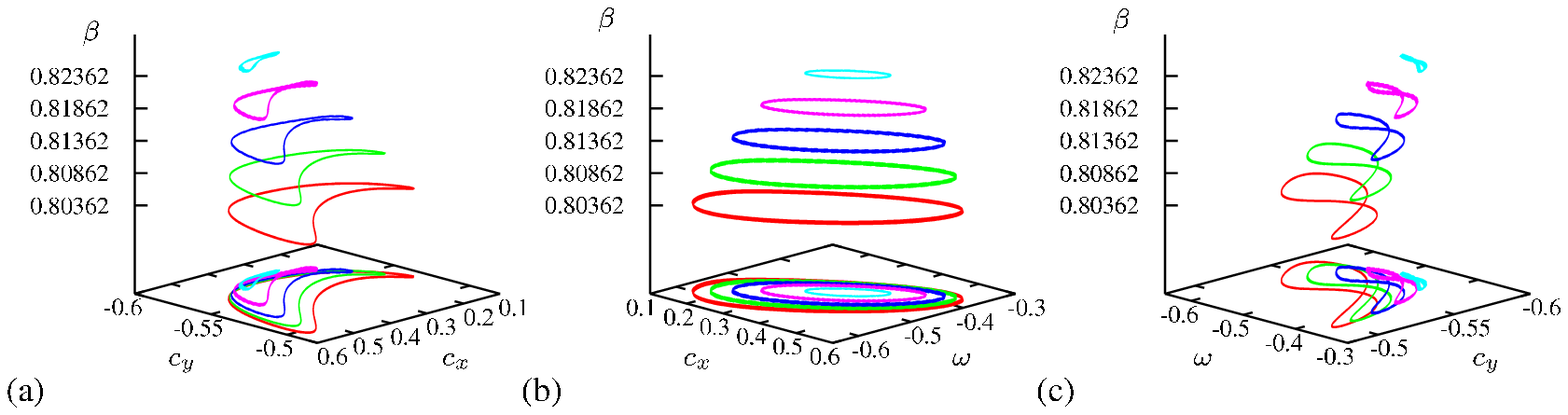}{
  (Color online)
  Same as \fig{eps02-quot}, for $\alpha=0.25$. 
}{eps025-quot}

The parametric line $\alpha=0.25$ exhibits similar behaviour, as shown
in \fig{eps025-tip} and \figref{eps025-quot}. This is closer to the
Hopf bifurcation line in the quotient system, called $\dM$ line in
\cite{Winfree-1991}. Correspondingly, the size of the limit cycles in
the quotient system is smaller and they become more oval-shaped
Note that the scale of $c_y$-axis is disproportionately stretched in
\fig{eps025-quot}, \ie\ the Hopf central manifold appears to be 
nearly orthogonal to that axis.
Again, there is no qualitative change in the quotient system 
dynamics when crossing the 1:1 resonance.


\section{Application II: Large Core Spirals}\seclabel{largecore}

Another example where the spatial extent of the spiral wave dynamics
is large is the vicinity of Winfree's ``rotors boundary'' $\dR$
in the parametric space~\cite{Winfree-1991}. In the vicinity of this
boundary, the period of rotation and the radius of the core of the
spiral wave grow infinitely.

There are at least two different asymptotic theories, based on
different choice of small parameters, which aim to describe the
vicinity of $\dR$. Hakim and
Karma~\cite{Hakim-Karma-1997,Hakim-Karma-1999} have developed a
``free-boundary'' asymptotic theory applicable to FitzHugh-Nagumo type
models in the limit $c\to0$ or $\alpha\to0$ in terms of our chosen
kinetics, where angular velocity $\omega$ typically decreases as
\begin{equation}
  |\omega| \propto |p-p_*|^{\frac{3}{2}}, \qquad p\to p_*, \eqlabel{HKlaw}
\end{equation}
where $p$ is a parameter of the model such that $p=p_*$ corresponds to
the $\dR$ boundary.

Elkin~\etal~\cite{Elkin-etal-1998} obtained an alternative
asymptotic based on assumptions which were not restricted to
kinetics of any particular kind, but which were not directly
validated. Their prediction was
\begin{equation}
  |\omega| \propto |p-p_*|, \qquad p\to p_*. \eqlabel{EBHlaw}
\end{equation}

Further analysis has suggested that these two alternatives are not
actually antagonistic and may be even observed in the same system in
different parametric regions~\cite{Elkin-etal-2002}. Reliably
distinguishing between the two asymptotics is challenging for DNS as
it requires a rather close approach to the critical point $p=p_*$,
which is not known \apriori, implying large tip trajectory radii and
correspondingly significant computational resources.

In here we present an example of studying this dependence using
calculations in the comoving FoR, which is free from the above
complication, as it can be performed within the box of fixed size for
all $p$.

For this study, we use Barkley's model with varying parameter $p$
chosen to be $a$, varying from $a=0.48$ downwards with step $0.001$ 
until $0.43$, with other parameters fixed at $b=0.05$ and
$c=0.02$. The discretization parameters are $\L=30$, $\stepx=1/8$,
$\stept=1/2560$ and $\rinc=(0,7/4)$.

\sglfigure{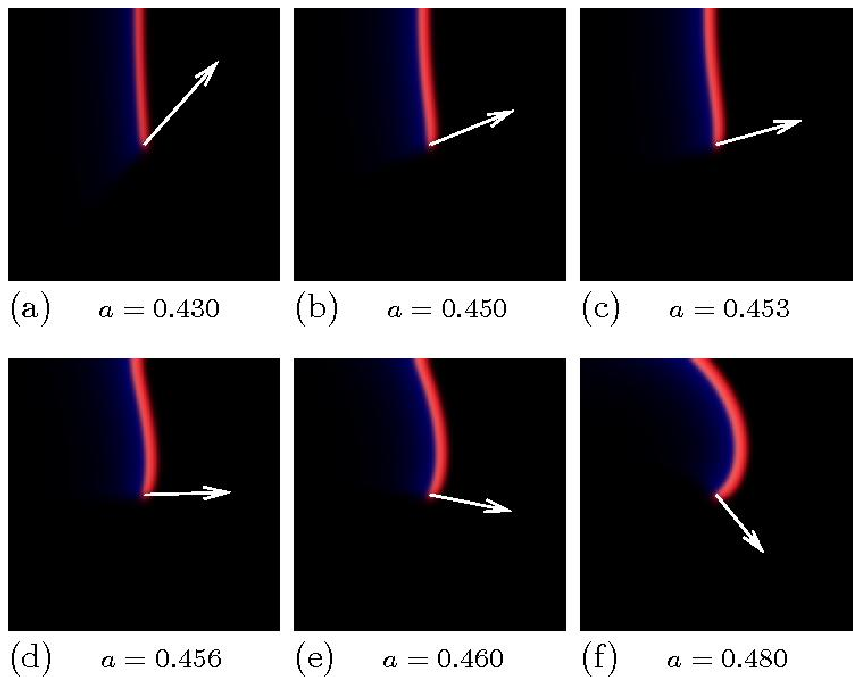}{
  (Color online)
  Snapshots of relative equilibria in Barkley model obtained at different values of parameter $a$.
  The arrows indicate the direction of the vector $\c$. 
}{lc-snaps}

\sglfigure{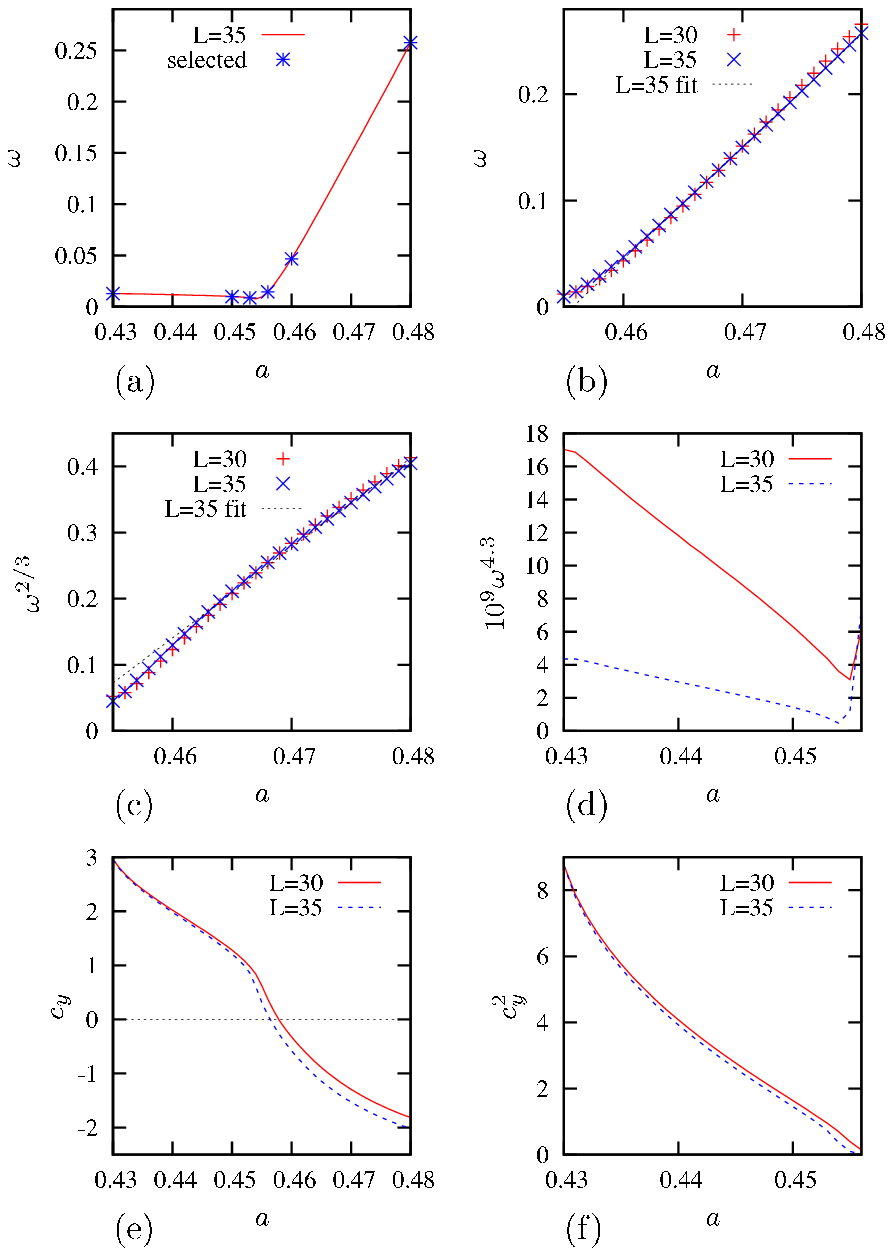}{
  (Color online)
  Dependencies $\omega(a)$ and $c_y(a)$ of the relative equilibria, for different $L$ as indicated. 
  On panel (a), the symbols correspond to the selected values of $a$ used in \fig{lc-snaps}. 
}{lc-curves}

Selected stationary solution obtained in this way are illustrated in
\fig{lc-snaps}, and the graphs of $\omega(a)$ and $c_y(a)$ are shown
in \fig{lc-curves}. We compare the features of the observed
solutions with those that are given by the two asymptotic
theories~\cite{Elkin-etal-2002}, and observe that
\begin{enumerate}
\item\label{finger} There is a critical value of the parameter $a_*\approx 0.456$,
  at which the behaviour of the solution changes qualitatively. At
  $a=a_*$, we observe a nearly straight broken excitation wave. 
\item\label{rotating} For $a>a_*$, the solutions are spiral waves,
  that is, broken excitation wavelets, which become less and less
  convex as $a\to a_*$, and have macroscopic angular velocity which
  however diminishes in the same limit.
\item\label{translating} For $a<a_*$ the solutions are retracting
  nearly straight but slightly concave wavelets, with very small
  angular velocity.
\item\label{fingerrate} For $a=a_*$, the direction of movement of the
  tip seems approximately orthogonal to the overall orientation of
  the wave itself.
\item\label{retraction-rate} For $a<a_*$, the vertical component of 
  vector $\c$ depends on $a$ in a way which is consistent with the
  asymptotic $|c_y|\propto|a-a_*|^{1/2}$, see \fig{lc-curves}(e,f). Since
  the overall orientation
  of the wavelets, as seen in \fig{lc-snaps}(a--c), is nearly vertical we can
  take $c_y$ as a crude estimate
  of the ``global tip growth rate'' as defined in
  \cite{Elkin-etal-2002}.
\item\label{rotation-velocity} For $a>a_*$, the angular velocity of
  solutions depends on $a$ in a way which is consistent with the
  asymptotic $|\omega|\propto|a-a_*|$, see \fig{lc-curves}(b) but not
  $|\omega|\propto|a-a_*|^{3/2}$, see~\fig{lc-curves}(c).
\end{enumerate}

All these observations are in agreement with the theory in
\cite{Elkin-etal-2002} and can be used to empirically distinguish
between the Elkin~\etal\ asymptotics (corresponding to the ``I/V''
parametric boundary in \cite{Elkin-etal-2002}) and Hakim-Karma
asymptotics (respectively, ``J/C'' boundary in
\cite{Elkin-etal-2002}).

Feature~\ref{finger} is inconclusive: existence of a critical
solution, called ``critical finger'' by Hakim and Karma, is common for
both J/C and I/V boundaries, but the shape of this solution is
different. It is asymptotically linear for I/V boundary, and
asymptotically logarithmic for J/C boundary. Looking at
\fig{lc-snaps}(d) and considering the effect of the boundary
conditions, it is not clear which case is nearer to the observed
reality.

Feature~\ref{rotating} is common for I/V and J/C boundaries. The
phenomenological difference is that spirals close to I/V boundary can
be ``growing'' or ``shrinking'', while spirals close to J/C boundary
can only be ``growing''. The movement of the tip in
\fig{lc-snaps}(d--f) seems approximately orthogonal to the orientation
of the wavelet near the tip, which is consistent with both cases.

Feature~\ref{translating} tips the balance in favour the I/V boundary
since the broken wavelets are concave. According to
\cite{Elkin-etal-2002}, the translating waves near an I/V boundary
should be concave, and those near an J/C boundary should be convex.

Feature~\ref{fingerrate} is common for I/V and J/C, as in both cases the
critical fingers should have zero ``global growth rate''.

Feature~\ref{retraction-rate} is common for I/V and J/C boundaries. 

Feature~\ref{rotation-velocity} is, in our opinion, a convincing evidence
in favour of an I/V boundary, since according to~\cite{Elkin-etal-2002},
near I/V boundary the dependence $\omega(\delta)$ is linear,
whereas near J/C boundary it is $|\omega(\delta)|\propto|\delta|^{3/2}$. 

An unequivocal interpretation of all theoretical predictions in the
view of our present numerical results would require further
investigation, as the asymptotics of
\cite{Elkin-etal-1998,Elkin-etal-2002} operate with a ``crest line''
of an excitation wave. There is no obvious operational definition of
this line which would be valid up to the tip, and some of the
predictions concern the mutual orientation of this line and the tip
velocity. However the predictions that can be unambiguously
interpreted, seem to indicate that for the model considered here, we
have the case of I/V boundary, \ie\ Elkin~\etal\ asymptotics, rather
than J/C boundary corresponding to Hakim-Karma asymptotics.

The last observation here is that of the small angular velocity
$\omega$ calculated for the ``retracting waves'' at $a<a_*$, seen on
\fig{lc-snaps}(a--c). As we already noted, the smallness of these $\omega$ values
is consistent with the theoretical
prediction of translating but not rotating waves. However when these
values are magnified, we observe that they demonstrate a peculiar
power law $|\omega(a)|\propto|a-a_*|^p$ where $p\approx1/4.3$, see
\fig{lc-curves}(d). A theoretical explanation of this requires further
study; it is clear, however, that $\omega$ in this area is strongly
affected by the boundaries, as the curves for $\L=30$ and $\L=35$
differ quite significantly.

\section{Application III: Electrophoresis of meandering spiral}\seclabel{drift}

Finally, we illustrate calculation of the movement of spiral waves in
a perturbed reaction-diffusion system. We consider FitzHugh-Nagumo
kinetics at the same parameters as in \fig{fhn-movie}, and add to it the
``electrophoresis'' perturbation \eq{perturbation} in the right-hand side,
with $\E=\epsilon\D$. 

\dblfigure{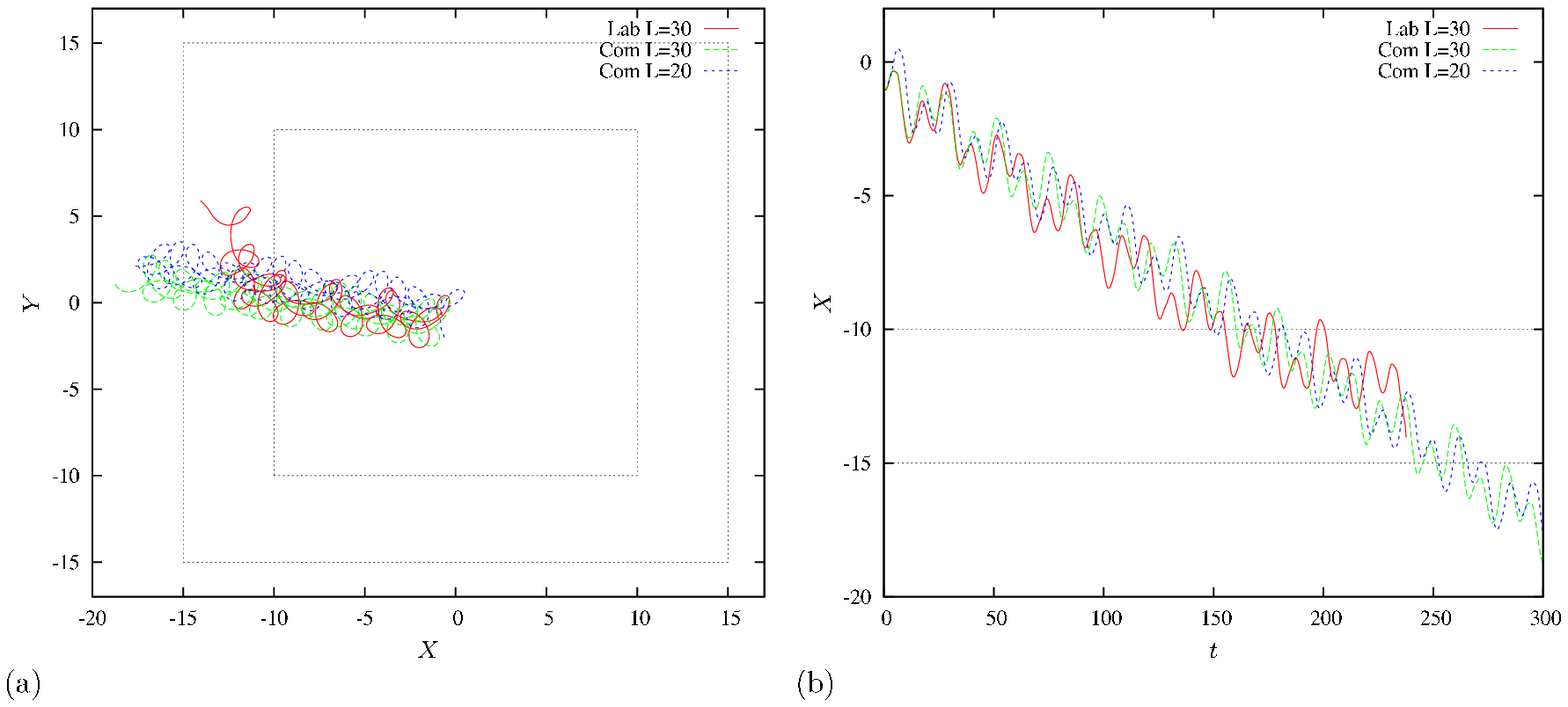}{
  (Color online)
  Trajectories of tips of drifting meandering spirals calculated in
  the laboratory FoR (for $\L=30$) and in the comoving FoR (for
  $\L=30$ and $\L=20$). The thin black dotted lines designate the
  boundaries of the calculation box in the laboratory FoR where the
  initial position of the tip is in the centre. The parameters are the
  same as in \fig{fhn-movie} and the perturbation is
  $\h=\epsilon\D\partial_x\u$, where $\epsilon=0.1$.
}{fhndr}

Results of the simulations are presented in \fig{fhndr}. The
unpertubed spiral waves for these parameters are meandering, so with
the perturbation present, we observe meandering with drift. The drift
proceeds with a constant average velocity, which is consistent with
the fact that the perturbation violates only the rotational but not
the translational symmetry of the problem. The average drift is to the
left, which corresponds to a collapsing scroll wave. So at these
parameter values, the scroll waves have positive tension, inasmuch as
this concept can be applied to meandering scrolls. 

In the calculations in the laboratory FoR, the time during which the
drift can be observed is limited, as when the spiral reaches the left
boundary, it terminates. In the comoving FoR, this drift can be
observed indefinitely. Comparing the traces in \fig{fhndr} we see that
although, as we know from \figs{clips} and
\figref{mean}, the discretization is too crude to give
quantitative agreement between laboratory and comoving calculations in
detail, the drift velocities obtained in these two ways are very
similar. 

We illustrate the relative advantages of the two methods of
calculation by comparing the computation costs. The laboratory FoR
simulation, for $\L=30$ and $t\in[0,300]$ has taken 325~sec (the
spiral has annihilated at the left at $t\approx237$). The time taken by
the comoving FoR simulation for the same boxsize $\L$ and the same $t$
interval is 462~sec, \ie\ is naturally somewhat longer due to the
extra effort required for the advection term calculations. However,
the comoving FoR calculation proceeded unabated where the laboratory
FoR calculation failed due to annihilation with the border. To
continue the laboratory FoR calculation to the same extent we would
have to increase the box size $\L$ with a corresponding increase in
computation cost. Moreover, virtually the same result, as far as drift
velocity is concerned, can be obtained by comoving FoR calculation
with $\L=20$, and it takes only 202~sec. Of course the drift in the
laboratory FoR with $\L=20$ would terminate even earlier.


\section{Discussion}\seclabel{discussion}

We have described a numerical method of solving a reaction-diffusion
system of equations describing a spiral wave, in a frame of reference
which is moving with the tip of that wave. 

We have shown the method can provide accurate solutions, and that
there are applications where the computational cost of our method can
be considerably lower than that of the conventional approach, or
the conventional approach is just inapplicable.~
  As always, the computational advantages are particularly essential
  in case of parametric studies, for which the method is well suited. 

Although the applications were chosen just to provide some meaningful
examples of use of the method, the results obtained there can
be of scientific value themselves. 

So, we have investigated the vicinity of the ``1:1 resonance''
manifold in the parametric space, which corresponds to spontaneous
drift of spirals, and which separates meandering patterns
with outward petals and inward petals.
Henry~\cite{Henry-2004} has proposed a theory
which implies that this
manifold coincides or is an analytical continuation of the manifold
where the filament tension of scroll waves vanishes. 
There are reports in literature confirming that change of sign
of filament tension is
associated with change from outward to inward petals in meandering
patterns, but also examples where there are no such 
correlation, \eg~\cite{Alonso-Panfilov-2008} and references therein.
Our
simulations indicate that as far as orbit manifold dynamics of the
spiral is concerned, the 1:1 resonance is not characterized by any
special features. Hence any special features of this resonance ought
to be due to the Euclidean extension of the orbifold dynamics. Since
scroll filament tension can also be defined via properties of the
spiral wave solutions within the comoving FoR, any
genetic and generic relationship between the two manifolds seems
unlikely (but, of course, cannot exclude the possibility of such
relationship in some special cases).

We have also investigated the vicinity of the ``$\dR$'' manifold in
the parametric space, which has provided a strong evidence towards one
of the two theoretical possible asymptotics, namely
Elkin~\etal~\cite{Elkin-etal-1998} asymptotics as opposed to
Hakim-Karma~\cite{Hakim-Karma-1997} asymptotics. It should be noted
here that while Hakim-Karma asymptotic theory was based on assumptions
which have been well established, the Elkin~\etal\ asymptotic theory
was using assumptions, validity of which could not be asserted at that
moment. 
Here we have presented firm evidence that
Elkin~\etal\ asymptotis is not a mere theoretical possibility but is
indeed observed in reality
(see also~\cite{Hermann-Gottwald-2010} and a discussion below).
A direct confirmation would be via
calculation of the ``response functions'', \ie\ critical
eigenfunctions of the adjoint linearized operator of the critical
finger solution. This would require obtaining first a good quality
critical finger solution, so the method described here can be a
significant step towards this goal, too.

Finally, we have demonstrated that calculations in the comoving FoR
can be efficiently used to study perturbation-caused drift of spirals,
including meandering spirals. Although the asymptotic theory of drift
of meandering spirals is yet to be developed (see, however, a
preliminary draft of such theory in~\cite{Foulkes-2009}), we can
expect, for instance, that scroll waves in the FitzHugh-Nagumo model
with the parameters as in \fig{fhn-movie}, \figref{fhndr} will have ``positive
tension'', \ie\ tend to collapse, rather than develop a scroll wave
turbulence. The advantage of calculating drift in the comoving FoR,
apart from computation cost, is absence of ``pinning'' effects of
spatial discretization, both in terms of discrete space steps and
discrete spatial directions, on the drift.

 Our approach can be compared to the approach proposed by Beyn
  and Thummler (BT)~\cite{Beyn-Thummler-2004}. BT use a similar
  mathematical idea of decomposing the evolution of the nonlinear wave
  into the motion of the wave and evolution of its shape, which in the
  functional space appears as decomposition into motion along and
  across the Euclidean symmetry group orbits. But there are also
  differences.  There are technical details of implementations which
  are probably of lesser importance, such as choice of polar vs
  Cartesian grid, central vs upwind discretization of spatial
  derivatives and explicit vs semi-implicit discretization in time.
  More significant differences are in the ``phase conditions'' they
  use, which play the same role as, but are qualitatively different in
  nature from, our ``pinning conditions''. One aspect is that the
  phase conditions involve integral functionals. We show here that
  this is not necessary, and local conditions like~\eq{musteps} are
  simpler. The other aspect is the one we discussed in \secn{maths}:
  the BT phase conditions appear to be well suited for
  calculation of relative equilibria (rigidly rotating spirals) but
  not necessarily for relative periodic solutions (meandering
  spirals). Further, the phase conditions proposed by BT were not
  intended for use with symmetry breaking perturbations that produce
  drift of spirals. And indeed, BT comment in their paper that ``it
  seems quite a challenging task to freeze drifting spirals or
  recognize meandering spirals as periodic orbits.'' As we have
  demonstrated, our approach works both for meandering spirals and for
  drifting spirals.

  After completing this study we became aware of a work by Hermann and
  Gottwald (HG) \cite{Hermann-Gottwald-2010} who also investigated the
  large core limit, using a further development of the BT method. HG have
  paid a great deal of attention to refining the boundary conditions
  so as to minimize the effect of boundaries onto the quotient
  dynamics. This has allowed them, in particular, to verify the linear
  scaling law \eq{EBHlaw} for seven decades of variation of
  $|\omega|$, compared to mere one decade as shown
  in~\fig{lc-curves}. Notice that as shown in the same figure, our
  progress towards smaller values of $|\omega|$ is limited precisely
  by the influence of boundaries.  HG also have explicitly addressed
  the issue of the numerical stability of the computations, which we
  treat in this study purely empirically.

  We believe that combining the advantageous features of the approach
  developed by BT and HG, and the one proposed here, is an interesting
  topic for future work, which may yield further results about spiral
  wave dynamics, that are not possible, or very difficult, to obtain
  by direct numerical simulations.  


\section*{Acknowledgements}
This study has been supported in part by a Ph.D studentship from the
University of Liverpool and by EPSRC grant~EP/D074789/1.
The preliminary stage for this study was done in
  collaboration with B.N.~Vasiev, who proposed the idea of solving for
  quotient data simultaneously with the advection substep.  VNB is
  also grateful to A.M.~Pertsov and G.A.~Gottwald for inspiring
  discussions, and to G.A.~Gottwald also for informing us about the
  results of~\cite{Hermann-Gottwald-2010} prior to publication.


\end{document}